\documentclass[iop,revtex4]{emulateapj}
\usepackage{graphicx}
\usepackage{amssymb}
\usepackage{threeparttable}
\usepackage{lscape}
\usepackage{hhline}

\mathchardef\mhyphen="2D
\newcommand{\aliii}{Al\,{\sc iii}}

\newcommand{\ovi}{O\,{\sc vi}}

\newcommand{\feii}{Fe\,{\sc ii}}

\newcommand{\siv}{S\,{\sc iv}}
\newcommand{\siiv}{Si\,{\sc iv}}
\newcommand{\siii}{Si\,{\sc ii}}

\newcommand{\cii}{[C\,{\sc ii}]}
\newcommand{\ciii}{C\,{\sc iii}}
\newcommand{\civ}{C\,{\sc iv}}

\mathchardef\mhyphen="2D

\def\cii{C\,{\sc ii}}
\def\ciii{C\,{\sc iii}}
\def\civ{C\,{\sc iv}}
\def\niii{N\,{\sc iii}}

\def\oiv{O\,{\sc iv}}
\def\ovi{O\,{\sc vi}}

\def\nevi{Ne\,{\sc vi}}

\def\siiv{Si\,{\sc iv}}
\def\Siii{Si\,{\sc ii}}
\def\siii{S\,{\sc iii}}

\def\siv{S\,{\sc iv}}

\def\aliii{Al\,{\sc iii}}

\def\feii{Fe\,{\sc ii}}

\def\nh{\ifmmode n_\mathrm{\scriptstyle H} \else $n_\mathrm{\scriptstyle H}$\fi}
\def\ne{\ifmmode n_\mathrm{\scriptstyle e} \else $n_\mathrm{\scriptstyle e}$\fi}
\def\qh{\ifmmode Q_\mathrm{\scriptstyle H} \else $Q_\mathrm{\scriptstyle H}$\fi}
\def\uh{\ifmmode U_\mathrm{\scriptstyle H} \else $U_\mathrm{\scriptstyle H}$\fi}
\def\Nh{\ifmmode N_\mathrm{\scriptstyle H} \else $N_\mathrm{\scriptstyle H}$\fi}

\def\Zsun{\ifmmode {\rm Z}_{\odot} \else Z$_{\odot}$\fi}
\def\Msun{\ifmmode {\rm M}_{\odot} \else M$_{\odot}$\fi}
\def\kms{\ifmmode {\rm km~s}^{-1} \else km~s$^{-1}$\fi}
\def\Lya{\ifmmode {\rm Ly}\alpha \else Ly$\alpha$\fi}
\def\Lyb{\ifmmode {\rm Ly}\beta \else Ly$\beta$\fi}
\def\Lyg{\ifmmode {\rm Ly}\gamma \else Ly$\gamma$\fi}
\def\Lyd{\ifmmode {\rm Ly}\delta \else Ly$\delta$\fi}
\def\neaod{\ifmmode n_\mathrm{\scriptscriptstyle AOD} \else $n_\mathrm{\scriptscriptstyle AOD}$\fi}
\def\necrit{\ifmmode n_\mathrm{\scriptstyle cr} \else $n_\mathrm{\scriptstyle cr}$\fi}
\def\ncr{\ifmmode n_\mathrm{\scriptstyle cr} \else $n_\mathrm{\scriptstyle cr}$\fi}
\def\nepi{\ifmmode n_\mathrm{\scriptscriptstyle PI} \else $n_\mathrm{\scriptscriptstyle PI}$\fi}
\def\gtorder{\mathrel{\raise.3ex\hbox{$>$}\mkern-14mu\lower0.6ex\hbox{$\sim$}}}
\def\ltorder{\mathrel{\raise.3ex\hbox{$<$}\mkern-14mu\lower0.6ex\hbox{$\sim$}}}
\newcommand{\vy}[2]{#1_{\scriptscriptstyle #2}}

\shorttitle{ }
\shortauthors{}

\begin{document}


\title{Evidence that 50\% of BALQSO outflows are situated at least 100 pc from the central source}


\author{
Nahum Arav\altaffilmark{1,$\dagger$},
Guilin Liu\altaffilmark{1}, 
Xinfeng Xu\altaffilmark{1},
James Stidham\altaffilmark{1},
Chris Benn\altaffilmark{2}
and
Carter Chamberlain\altaffilmark{1}
}

\affil{$^1$Department of Physics, Virginia Tech, Blacksburg, VA 24061, USA}
\affil{$^2$Isaac Newton Group, Apartado 321, E-38700 Santa Cruz de La Palma, Spain}

\altaffiltext{$\dagger$}{Email: arav@vt.edu}


\begin{abstract}

The most robust way for determining the distance of quasar absorption outflows is the use of troughs 
from ionic excited states. The column density ratio between the excited and resonance states yields the outflow number density.  Combined with a knowledge of the outflow's ionization parameter, a distance from the central source (R) can be determined. Here we report results from two surveys targeting outflows that show troughs  from \siv. One survey includes 1091 SDSS and BOSS quasar spectra, and the other includes higher-quality spectra of 13 quasars observed with the Very Large Telescope. Our \siv\ samples include 38 broad absorption line (BAL) outflows, and four mini-BAL outflows. 
The \siv\ is formed in the same physical region 
of the outflow as the canonical outflow-identifying species \civ. 
Our results show that \siv\ absorption is only detected in 25\% of \civ\ BAL outflows.  
The smaller detection fraction is due to the higher total column density ($N_H$) needed to detect \siv\ absorption. Since R empirically anti-correlates with $N_H$ 
the results of these surveys can be extrapolated to  \civ\ quasar outflows with lower $N_H$ as well.
We find that at least 50\% of quasar outflows 
are at distances larger than 100 pc from the central source, and at least 12\% are at distances 
larger than 1000 pc. These results have profound implications to the study of the origin and 
acceleration mechanism of quasar outflows and their effects on the host galaxy.

\end{abstract}






\clearpage

\section{INTRODUCTION}
\label{sec:intro}

Roughly one-third of observed quasars show outflows as blueshifted absorption troughs with respect 
to the active galactic nucleus (AGN) rest-frame spectrum \citep{Hewett03,Dai08,Dai12,Ganguly08,Knigge08}. 
Theoretical modeling has shown that such outflows can be a main agent of the so-called quasar-phase AGN feedback
\citep[e.g.][]{Ostriker10,Ciotti10,Soker10,McCarthy10,Hopkins10,FaucherGiguere12,Choi14}. 
However, in order to gain insight on the nature of these outflows and their influence on the 
host galaxy, it is critical to constrain their distance  from the central source ($R$). 
For example, most theoretical models predict that luminous quasars' outflows are accretion disk winds, which are observed in their 
acceleration phase at $R\sim0.01$ pc \citep[e.g.][see \S~7.3 for elaboration]{Murray95,Elvis00,Proga00}.

The most reliable observational method for determining $R$ in AGN absorption outflows relies on measuring 
troughs from excited states of various ions. In \S~\ref{sec_distance_measuring} we describe this method in detail, and in 
\S~\ref{sec:discussion} we compare this method with other empirical methods for determining $R$. Using the excited-state method
over the last decade, our group and others measured $R$ for about 20 AGN outflows, as well as their mass flow 
rate and kinetic luminosity \citep{Hamann01,deKool01,deKool02a,deKool02b,Gabel05,Moe09,Bautista10,
Dunn10,Aoki11,Arav12,Borguet12a,Borguet12b,Borguet13,Edmonds11,Arav13,Lucy14,Finn14,Chamberlain15a,Chamberlain15b}. 
All of these investigations located the outflows at $R$ of several pc to many kpc. For luminous quasars, 
the majority of these findings were at $R$ of hundreds to thousands of pc scales (see \citealt{Arav13} for 
a review). 

The majority of the $R$ determinations referenced above arise from singly ionized species
(e.g., \feii\ and \Siii). However, most outflows show absorption troughs only from more highly ionized species. 
Therefore, the applicability of $R$ derived from singly ionized species to the majority of outflows is somewhat 
model-dependent (see discussion in \S~1 of \citealt{Dunn12}). To address this issue, several 
$R$ determinations were done using doubly and triply ionized species. 
Three of these studies used HST to observe transitions from excited states shortward of 1000\AA. 
These published results all found $R>500$~pc  
(3000 pc using \oiv\ and \oiv*, \citealt{Arav13}; 4000 pc using \oiv\ and \oiv*, \citealt{Finn14}; 
800 pc using \niii\ and \niii* \citealt{Chamberlain15b}), strongly suggestive of a large size scale for quasar outflows.
However, the paucity of HST observations does not allow for a robust statistical conclusion regarding the 
distribution of $R$ values. 

In contrast, ground-based observations have provided many thousands of spectra of quasar outflows.
From the ground, the main high-ionization species with a measurable trough arising from an excited state is \siv,
which has resonance and excited level transitions at 1063 and 1072\AA, respectively (see \S~\ref{sec_distance_measuring}).  
It is difficult to obtain high-quality spectra of similar diagnostics shortward of 1000\AA, from the ground, due 
to the strongly increasing density of the \Lya\ forest at the redshifts needed to cover this spectral band. Therefore, $R$ determinations for high-luminosity quasars, using high-ionization diagnostics from the 
ground, concentrated on using the above \siv\ diagnostics. Four such analyses are found in the literature. In three 
of these, \siv\ yielded the following distances: $R>3000$ pc \citep{Borguet13}, $R=300$ pc \citep{Borguet13}, and 
$R=100$ pc \citep{Chamberlain15a}. In one outflow, the \siv\ diagnostics could not provide an accurate constraint, 
but by using \ciii* diagnostics, a distance range of 10 pc $<R<$ 300 pc was established \citep{Borguet13}.

However, analysis of individual objects suffers from selection biases. In order to establish the distribution of 
$R$ in the general population of quasar outflows using the \siv\ diagnostics, we need to analyze unbiased samples --- unbiased in the sense that the depth ratio of the \siv\ and \siv* troughs is not known a priori (see elaboration in \S~\ref{sec_distance_measuring}).
In this paper, we present the analysis results of two such samples. The first is comprised of 1091 SDSS quasar 
spectra, and the second contains a blind sample of 13 outflows obtained with the Very Large Telescope (VLT) X-shooter spectrograph that 
have much higher data quality than the SDSS observations.
 
This paper is organized as follows. In \S~\ref{sec_distance_measuring}, we describe the physics and observational 
framework that allow us to use the \siv\ troughs to determine $R$. In \S~\ref{sec:survey}, we describe the samples selection, 
and in \S~\ref{sec:Identification}, we describe the process of identifying and measuring the \siv\ and \siv* troughs. We present 
our results in \S~\ref{sec:results}. 
In \S~\ref{sec:extrapolation} we detail the steps and assumptions that are needed to extrapolate these results to the general population of high ionization BALs (HiBAL) outflows.
 In \S~\ref{sec:discussion},  we discuss these results, including their extrapolation for general broad absorption line (BAL) outflows, and show that the criticism about using  the \siv\ diagnostics for $R$ determination \citep{Lucy14} is unwarranted. We summarize this work in \S~\ref{sec:summary}.

\section{Using \siv\ and \siv* Troughs to Measure Distances of AGN Outflows}\label{sec_distance_measuring}

As discussed in \citet{Borguet12b}, the \siv\ multiplet around 1070\AA\ is composed of three lines: the 
ground-state transition with a wavelength of 1062.66\AA, and two transitions arising from an excited state 
($E=951$ cm$^{-1}$, hereafter, referred to as \siv*) at wavelengths 1072.96\AA\ and 1073.51\AA\, where the excited
state being populated at a critical density of $\ncr=5.6\times10^4$~cm$^{-3}$ (at 10,000 K).
Due to their small separation, for most applications we combine the excited 1072.96\AA\ and 1073.51\AA\
transitions to one transition (but see important exception in \citealt{Borguet12b}, where the two excited-state 
troughs are somewhat resolved). The oscillator strength ($f$) of the combined excited transitions is equal 
to the $f=0.05$ of the 1062.66\AA\ resonance transition \citep{Verner96}. In passing, we mention that due to the same number 
of electrons in the outer shell, the following ions have a similar multiplet transition configuration that 
can also be used to extract measurements of electron number density (\ne) in AGN outflows: 
\cii~\citep[e.g.][]{Edmonds11}, \niii~\citep[e.g.][]{Chamberlain15b}, \oiv~\citep[e.g.][]{Arav13}, and 
\nevi\ and \Siii~\citep[e.g.][]{Borguet12b}. The method described below for \siv\ can be applied to the above ions as well.

The \siv* energy level is predominantly populated via collisions of free electrons with \siv\ in the ground state \citep[e.g.,][]{Leighly09}. Thus, the ratio of the column densities of \siv\ and \siv*, $N$(\siv)/$N$(\siv*), 
can be used as an \ne\ diagnostic \citep[e.g.][equations (3.20) and (3.23)]{Osterbrock06}:

\begin{equation}
\ne\simeq \ncr\left[\frac{2\,N(\mbox{\siv})}{N(\mbox{\siv*})}e^{-\Delta E/kT}-1\right]^{-1},
\label{eqn:ne}
\end{equation}
where \ncr\ is the critical density for the excited and resonance energy levels of \siv, $\Delta E$ is the 
energy difference between the ground- and excited-state levels, $k$ is the Boltzmann constant, and $T$ is the temperature in Kelvin.  
The factor 2 comes from the ratio of degeneracies for the \siv*\ and \siv\ energy level, respectively. For 
plasma photoionized by a quasar spectrum, the region where \siv\ exists has $T\simeq10^4$ K (see \S~\ref{sec:Temperature}),
for which $e^{-\Delta E/kT}=0.9$. For simplicity, for the discussion in this section, we assume $e^{-\Delta E/kT}=1$, and 
in section \S~7.4 we elaborate on the minor effect that 
a finite temperature will have on our results.

In practice, we deduce the \ne\ of the outflow by comparing the $N$(\siv*)/$N$(\siv) ratio to predictions 
(see Fig.~1) made with ionic collision models that use the Chianti 8.0.2 atomic database \citep{Dere97,DelZanna15}. 
Such a prediction is shown in the right panel of figure~\ref{fig:siv_ap}. 

A photoionized plasma is characterized by the ionization parameter
\begin{equation}
U_\mathrm{H}\equiv\frac{{\displaystyle Q_\mathrm{H}}}{{\displaystyle 4\pi R^2 \vy{n}{\mathrm{H}} c}},
\label{eqn:U}
\end{equation}
where $Q_\mathrm{H}$ is the rate of hydrogen-ionizing photons emitted by the object, $c$ is the speed of light, $R$ 
is the distance from the central source to the absorber, and $\vy{n}{\mathrm{H}}$ is the total hydrogen number density.  
Using photoionization models, we can solve for \uh\ of a given quasar outflow \citep[e.g.][]{Arav01,Hamann01,Edmonds11,Chamberlain15a}. 
Furthermore, in the outflow zone where \siv\ is abundant, the plasma is highly ionized, and $\ne\simeq1.2\,\vy{n}{\mathrm{H}}$. 
Therefore, we can use equation (\ref{eqn:U}) to solve for $R$. This is the procedure used for deriving $R$ in most published 
results to date.

\subsection{The Ideal Apparent Optical Depth Case}

As described above, the ratio of occupied excited states to occupied ground states is a sensitive diagnostic for  \ne\ 
(see Fig.~\ref{fig:siv_ap}) and thus $R$. For the apparent optical depth (AOD) case (where the residual  intensity 
$I=e^{-\tau}$), $N$(\siv) and $N$(\siv*) can be straightforwardly extracted from the troughs  \citep[e.g., see equation (1) in ][]{Arav03}. Their ratio is then plotted 
on the theoretical curve (see Fig.~\ref{fig:siv_ap}), and \ne\ can be readily determined. 
There are two complications 
to address. First, there is contamination from \Lya\ forest absorption features, which is best addressed by using smooth templates 
to fit the \siv\ and \siv* troughs \citep[see][]{Borguet12b,Borguet13,Chamberlain15a}. The smooth \siv\ and \siv* optical depth templates used in the references above had the same functional form (as a function of velocity) up to a scalar scale factor (SF). The templates gave good fits for the troughs in these high-quality X-shooter data, and therefore motivated us to use a similar approach in this study as well.

The second complication is that at high \ne\ values, the ratio 
$N$(\siv*)/$N$(\siv) approaches an asymptotic value (the Boltzmann ratio), and thus small errors in $N$(\siv*) and $N$(\siv) will 
create large uncertainties in the \ne\ estimate. In the right panel of figure~\ref{fig:siv_ap}, we conceptually demonstrate this issue.  
The error bars on \ne\ are derived by the crossing point of the $\pm$ errors with the theoretical ratio curve. Therefore, the high-ratio case has a strong asymmetric uncertainties in \ne.

\begin{figure*}
\centering
\includegraphics[angle=90,scale=0.3,clip=false]{figSIV.ps}
\includegraphics[angle=90,scale=0.3,clip=false]{figSIV_linear.ps}
\caption{{\bf Left:} For a given \siv\ trough (blue solid line), there are three qualitatively different 
possibilities for the \siv* trough (dashed lines), which are discussed in \S~2.2 (the 0.5 and 1.5 values are only for illustrative purposes). 
{\bf Right:} For the AOD case, the column densities of the \siv\ and \siv* troughs can be directly measured, 
and their ratio yields \ne\ (see Equation \ref{eqn:ne}). This \ne\ value, combined with a determination  of \uh\ (see Equation \ref{eqn:U}), 
yields the distance of the outflow from the central source. For a typical \siv\ outflow in a luminous quasar, the resultant 
distance is displayed on the top x-axis (see \S~2.3).
}
\label{fig:siv_ap}
\end{figure*}

\subsection{The Realistic Nonblack Saturation Case}

Extracting reliable $N$(\siv) and $N$(\siv*) from an outflow's spectrum is complicated by the phenomenon of nonblack saturation. 
Using ionic doublet and multiplet troughs, it has been shown extensively that AOD-extracted $N_{ion}$ often underestimates the 
actual $N_{\rm ion}$, even in cases where the absorption features are fully resolved spectroscopically \citep[e.g.][]{Arav97,Arav99,Arav03,Arav08}, sometimes by as much as a factor of 1000 \citep{Borguet12b}.  
With the possibility of nonblack saturation, we need to carefully assess the information available about $N$(\siv) and $N$(\siv*) 
that is embedded in their troughs.

For resonance doublets (e.g., \civ~$\lambda\lambda$1548, 1551), which arise from the same energy level, the real optical depths of 
the troughs obey the ratio $\tau$(\civ 1548\AA)=2$\tau$(\civ 1551\AA) due to the 
ratio of their oscillator strengths. For \siv, the 1062 and 1073\AA\ transitions arise from separate  energy levels with 
different electron populations. Therefore, there is no a priori relationship between their real optical depths. However, as we show 
below, even after accounting for the possibility of nonblack saturation, we can extract definitive constraints about \ne\ from the 
observed depth ratio of the  \siv\ and \siv* troughs. 

We define $N$(\siv*)$_{\mathrm{AOD}}$ and $N$(\siv)$_{\mathrm{AOD}}$ as the column densities derived for each trough using the AOD 
method; we also define the value of \ne\ determined from the ratio of these two column densities as \neaod\ (see right panel of Fig.~\ref{fig:siv_ap}).
For a case where the \siv\ trough is deeper than the \siv* one, $N$(\siv)$_{\mathrm{AOD}}>N$(\siv*)$_{\mathrm{AOD}}$, and 
vice versa. Once we detect an \siv~$\lambda$1062 trough, there are three cases that cover all physical possibilities, detailed as 
follows.

\subsubsection{The \siv*~$\lambda$1073 trough Is shallower than the  \siv~$\lambda$1062 trough} 

To investigate the effects of saturation, we use the standard partial-covering model 
\citep[e.g., see \S~3.2, in][]{Edmonds11}, which 
assumes that an absorber with a single-value optical depth partially covers
a single homogeneous emission source. For this case, the normalized residual
intensity observed for the \siv\ and \siv* troughs ($I_{0}$ and $I_{*}$, respectively)  as a function of the
radial velocity $v$ can be expressed as \citep[see equation (1) in][]{Edmonds11}

\begin{equation}
 I_{0}(v)=1 - C(v) + C(v)e^{-\vy{\tau}{0}(v)}
\label{eq:pc0}
\end{equation}
\begin{equation}
 I_{*}(v)=1 - C(v) + C(v)e^{-\tau_{*}(v)}
\label{eq:pc}
\end{equation}

where $C(v)$ is the fraction of the emission source covered by the
absorber (we assume that $C(v)$ is the same for \siv\ and \siv* as they arise from the same ion). When the \siv\ trough is deeper than the \siv* one, $I_{0}(v) < I_{*}(v)$, and the physical range of the covering factor is $1-I_{0}(v)<C(v)<1$. For $C(v)=1$ we retrieve the AOD case $(\vy{\tau}{0}(v)=-\ln[I_{0}(v)]$ and $(\tau_{*}(v)=-\ln[I_{*}(v)])$. The case of maximum saturation is when $C(v)=1-I_{0}(v)$ for which equation 
(\ref{eq:pc0}) yields $\vy{\tau}{0}(v)\rightarrow \infty$. In contrast, for the same case, equation (\ref{eq:pc}) yields a finite $\tau_{*}(v)=-\ln[I_{*}(v)-I_{0}(v)]$.
For the example given by the solid and dashed blue lines in figure~\ref{fig:siv_ap}, at the bottom of the troughs, $\vy{\tau}{0}(v)_{\rm AOD}=1$,  while $\tau_{*}(v)_{\rm AOD}$=0.5.
For the same troughs, the case of maximum saturation yields $\tau_{*}(v)=0.97$, while 
$\vy{\tau}{0}(v)\rightarrow \infty$.

This example illustrates the general case that can be deduced straightforwardly from
equations (\ref{eq:pc0}) and (\ref{eq:pc}):
 
\begin{equation}
 \frac{\tau_{*}(v)}{\vy{\tau}{0}(v)}\leq \frac{\tau_{*}(v)_{\rm AOD}}{\vy{\tau}{0}(v)_{\rm AOD}}<1.
\label{eq:tauaod}
\end{equation}
The column density derived for each trough is proportional to the integrated optical depth over the width of the trough and inversely proportional to $f\lambda$, 
where $f$ and $\lambda$ are the oscillator strength and wavelength of the transition that gives rise to the trough. For the \siv\ and \siv* troughs we analyze, the $f$ values are the same, and the difference in $\lambda$ is less than 1\%. Combing this assertion with equation (\ref{eq:tauaod}), for the case where
$I_{0}(v) < I_{*}(v)$ we obtain
\begin{equation}
 \frac{N({\rm S\;{\scriptstyle IV}}^*)}{N({\rm S\;{\scriptstyle IV}})}\leqslant
\frac{N({\rm S\;{\scriptstyle IV}}^*)_{\rm AOD}}{N({\rm S\;{\scriptstyle IV}})_{\rm AOD}}<1,
\end{equation}
and therefore equation (1) yields $n_e\leq\neaod<\necrit$.

As discussed above, when the \siv* trough is shallower than the \siv\ trough,  only the \siv\ trough can be highly saturated, and therefore 
 the ratio $N$(\siv*)/$N$(\siv) can approach 0. So, in principle, \ne\ can be very small (see right panel in Fig.~\ref{fig:siv_ap}). However, the photoionization solution for the outflow using other measured ionic column densities (e.g., see Figures~6 and 7, and accompanying discussion in \citealt{Borguet12b})
will put an upper limit on $N$(\siv), which can then be used to derive $N$(\siv*) and obtain a lower limit for the $N$(\siv*)/$N$(\siv) ratio. 
We define the \ne\ associated with this limit as \nepi: the lower limit on \ne\ that arises from photoionization modeling.
The upper limit for the $N$(\siv*)/$N$(\siv) ratio is given by $N$(\siv*)$_{\mathrm{AOD}}$/$N$(\siv)$_{\mathrm{AOD}}$, 
and therefore the upper limit for \ne\ is \neaod. Thus, we obtain 
a range of possible \ne\ values $ \nepi<\ne<\neaod$ (shown as the shaded blue region in Figure 2).

An extreme version of this case is where an \siv*~$\lambda$1073 trough is not detected, and therefore only an upper limit $N$(\siv*)$_{\rm UL}$ can 
be obtained. For that case, 
\begin{equation}
 \frac{N({\rm S\;{\scriptstyle IV}}^*)}{N({\rm S\;{\scriptstyle IV}})}\leqslant
\frac{N({\rm S\;{\scriptstyle IV}}^*)_{\rm UL}}{N({\rm S\;{\scriptstyle IV}})_{\rm AOD}},
\end{equation}
and only an upper limit on \ne\ can be obtained.

It is straightforward to show that equations (5)-(7) also apply to inhomogeneous absorption models of outflows' troughs \citep[e.g.][]{deKool02b,Arav05,Arav08}.

\subsubsection{The \siv*~$\lambda$1073 Trough  Is Deeper than the  \siv~$\lambda$1062 Trough} 

An example of this case is shown by the red dashed line in the left panel of Fig.~\ref{fig:siv_ap}. For this case, 
\begin{equation}
 \frac{N({\rm S\;{\scriptstyle IV}}^*)}{N({\rm S\;{\scriptstyle IV}})}\geqslant
\frac{N({\rm S\;{\scriptstyle IV}}^*)_{\rm AOD}}{N({\rm S\;{\scriptstyle IV}})_{\rm AOD}}.
\end{equation}
This is again because the deeper trough is always more saturated than the shallower trough. However, there is an important physical difference 
in this case. The maximum value of $N$(\siv*)/$N$(\siv)=2, which is the ratio of degeneracies of the two energy levels. Moderate saturation can 
easily yield $N$(\siv*)/$N$(\siv)=2 for any case where the \siv*~$\lambda$1073 trough  is deeper than the  \siv~$\lambda$1062 trough. Therefore, in this case, $\necrit<\neaod<\ne$
 (see red arrow in Fig. \ref{fig:siv_sim}). 

\subsubsection{The \siv*~$\lambda$1073 Trough Is Equal in Depth to the  \siv~$\lambda$1062 Trough} 

\begin{figure}[t]
\centering
\includegraphics[angle=90,scale=0.29]{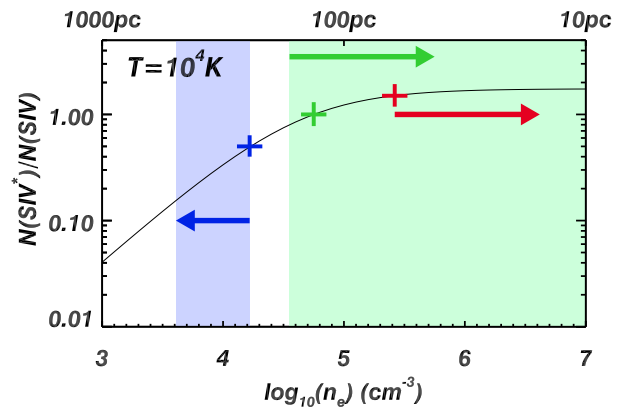}
\caption{Realistic nonblack saturation case (see \S~2.2). The ratio of \siv* to \siv\ column densities, 
$N$(\siv*)$_{\mathrm{AOD}}$/$N$(\siv)$_{\mathrm{AOD}}$, derived using the AOD method can be used to put important constraints 
on $R$ (see \S~2.3). When $N$(\siv*)$_{\mathrm{AOD}}$/$N$(\siv)$_{\mathrm{AOD}}<1$ (blue cross), the blue shaded region shows the allowed $R$ range. The AOD 
ratio sets the lower limit on $R$. If an \siv* trough is detected, an upper limit can be obtained from photoionization modeling. If a \siv* trough 
is not detected, only a lower limit on $R$ can be obtained using the upper limit for $N$(\siv*)$_{\mathrm{AOD}}$. 
If $N$(\siv*)$_{\mathrm{AOD}}$/$N$(\siv)$_{\mathrm{AOD}}=1$ (green cross), we can only set an upper limit on $R$ (leftmost part of the green shaded 
region), again using  photoionization modeling. When $N$(\siv*)$_{\mathrm{AOD}}$/$N$(\siv)$_{\mathrm{AOD}}>1$ the red cross yields the upper limit 
on $R$.} 
\label{fig:siv_sim}
\end{figure}

This occurrence is shown by the green dashed line in Fig.~\ref{fig:siv_ap}).
For this case, $N$(\siv*)/$N$(\siv) can be both larger and smaller than $N$(\siv*)$_{\mathrm{AOD}}$/$N$(\siv)$_{\mathrm{AOD}}$. This will depend 
on which trough is more saturated, which cannot be determined in this case. Thus, \ne\ could have any value above \necrit. The  lower limit for 
\ne\ is \nepi, as described in \S~2.2.1. The allowed region is shown as the green shaded region in figure \ref{fig:siv_sim}, and higher values of \ne\ to the right of the plot boundary are also allowed.

\begin{deluxetable*}{l c c}[ht]
\renewcommand{\arraystretch}{8}
\tablewidth{0.7\textwidth}
\tabletypesize{\small}
\setlength{\tabcolsep}{0.02in} 
\tablecaption{Distance from the central source based on an observed \siv~1062\AA\ Trough. \label{distance}}
\tablehead{
\colhead{\siv* 1073\AA\ Trough} &  \colhead{~\ne}\qquad\qquad & \colhead{Constraint on $R$ (pc)} 
}
\startdata
Undetected             &       $\;\;\;\;\;\;\;\;\;\;\,\ne<\neaod<\necrit$        &  $1000<R\quad\qquad\;\;\;$             \\
Shallower than \siv    &        $\nepi<\ne<\neaod<\necrit$                       &  $100<R<R_{\nepi}$   \\
Similar depth to \siv  &       $\nepi<\ne\qquad\qquad\quad\quad\;\,$             &  $\quad\quad\;\;\: R<R_{\nepi}$       \\
Deeper than \siv       &    $\necrit<\neaod<\ne\qquad\qquad\qquad\qquad\qquad\,$ &  $\quad\quad\, R<100$       \\    
\vspace{-2.2mm}
\enddata
\tablecomments{For the definitions of \neaod\ and \necrit\ see \S~2.2; for \nepi\ see the first paragraph of \S~2; and for $R_{\nepi}$ see \S~2.3.}
\end{deluxetable*}

\subsection{Distance Constraints from the \siv\ and \siv* Troughs Analysis}

In \S~2.2, we obtained constraints on \ne\ while taking into account the possibility of nonblack saturation in the \siv\ and \siv* troughs. 
We can convert these \ne\ constraints to constraints on $R$ as follows. For a given outflow, knowledge of \uh\ and \ne\ determines $R$ 
(see equation \ref{eqn:U} and accompanying discussion). However, with the limited signal-to-noise (S/N) ratio and spectral resolution of the SDSS data, it is difficult to derive reliable determinations of ionic column density from outflow troughs and therefore obtain \uh\ solutions for individual outflows. 
Also, the best we can do for constraining \ne\ is to classify the outflows into one of the three categories discussed in \S~2.2:  outflows that have $\ne<\ncr$ (\S~2.2.1), 
 outflows where $\ne>\ncr$ (\S~2.2.2), and outflows where \ne\ can be larger than \ncr\ (\S~2.2.3),

With these three cases in mind, and lacking individual $R$ determinations, we need to find a representative $R$ for \ne=\ncr\ and assess the plausible scatter around this value for individual 
objects. For high-luminosity quasars, there are four \siv\ outflows in the literature where detailed photoionization solutions are available 
(three can be found in table 5 of \citealt{Borguet13} and the fourth in table 5 of \citealt{Chamberlain15a}). The value of $R$ for each outflow 
was calculated using the \ne\ found for it. Therefore, using equation (\ref{eqn:U}), it is straightforward to derive $R$ for \ne=\ncr\ for each 
outflow. We find an average value of $R$(\ne=\ncr)=200 pc for these four outflows, where the individual values of $R$(\ne=\ncr) span the range 100--400 pc.
Recent results from the full analysis of our VLT/X-shooter sample (Xu et al., 2018, ApJ submitted) agree with the above $R$(\ne=\ncr) values (see elaboration in \S3.2).

We note that shielding-gas photoionization models (\citealt{Murray95}) yield much higher $U$ values for quasar outflows (and therefore smaller $R$ for a given $n_e$). However, the shielding models 
cannot reproduce the full set of extracted ionic column densities
from  high-quality observations of outflows 
 \citep[e.g.,][]{Hamann13}. Therefore, we make use of models that are able to adequately model observed outflows by reproducing the measured value of their trough's ionic column densities (i.e., \citealt{Borguet13,Chamberlain15a}, for actual \siv\ outflows).

The four objects mentioned above span magnitude and redshift ranges of
$18.36\geq r_{mag}\geq 17.65$ and $3.038\geq z\geq 2.076  $, respectively.
These ranges are similar to the ranges found for our SDSS and X-shooter samples,
$18.46\geq r_{mag}\geq 17.47$ and $3.180 \geq z\geq 2.004$.
Hence, the scatter of  
$R(\ne=\ncr)$ in our samples should be similar. Therefore, we expect that for the large majority of the \siv\ outflows in our samples, 
$R$(\ne=\ncr)$>$100 pc, and we use $R$(\ne=\ncr)=100 pc as a lower limit. In \S~5.3 we show that for cases where a \siv* trough is not detected
[$N$(\siv*)$_{\mathrm{AOD}}$/$N$(\siv)$_{\mathrm{AOD}}\ltorder0.1$], $R\gtorder1000$~pc.

In table \ref{distance} we summarize the \ne\ and $R$ constraints that can be deduced for a high-luminosity quasar where an \siv~1062\AA\ outflow 
trough is observed. We use the full constraints detailed in \S~2.2 and define $R_{\nepi}$ as the distance of the outflow when \ne=\nepi.

\section{\siv\ survey samples}
\label{sec:survey}

Using the SDSS data release 7, the final release of SDSS-II \citep{Schneider10}, \citet{Dunn12} compiled 
a sample of 156 bright quasars covering the \siv\ and \siv* spectral region. They found three quasar outflows that show both \siv\ and \siv* troughs (a fraction of 1.9\%). 
However, the \citet{Dunn12} survey does not yield information about the distribution of $R$ in quasar outflows. The number of \siv\ outflows (three) is too small for a census, and more importantly,  the criterion therein requiring 
the existence of both \siv\ and \siv*\ causes a bias against weak or nonexistent \siv*\ troughs, leading to 
uncertainty in the census of $R$ values. 
In this paper, we overcome these shortcomings by conducting an SDSS \siv\ survey using a sample eight times larger than that of 
\citet{Dunn12}, and requiring only an identification of the \siv\ trough.

We also analyze a sample of objects observed with the VLT/X-shooter spectrograph, which has the advantage of much better data than the SDSS spectra.

\subsection{The SDSS Sample}

The final released quasar catalog of SDSS-III has been made publicly available through the SDSS 
website\footnote{http://data.sdss3.org/datamodel/files/BOSS\_QSO/DR12Q/\\DR12Q.html}. This catalog
is a product of the intensive visual inspection of SDSS optical spectra from the twelfth data release
\citep[DR12;][]{Alam15} undertaken by \citet{Paris17}. Using this catalog,
we construct our sample by selecting all the quasars satisfying the following criteria.
\begin{enumerate}
 \item The $r$-band magnitude, obtained through point spread function (PSF) fitting, is required to be 
brighter than 18.5, which facilitates an S/N ratio high enough for performing absorption-line identification and kinematic analysis. 
 \item the spectral coverage of the SDSS and BOSS spectrographs are
3800--9100\AA\ and  3600--11000\AA, respectively. 
The redshift \citep[visually adjusted by][]{Paris17} is required to be higher than 2.8 for SDSS quasars and 2.6 for BOSS quasars.
At such redshifts,  we obtain spectral coverage down to 1000\AA\ in the rest frame for a BOSS $z=2.6$ and for an SDSS $z=2.8$ quasars, so the region
of an \siv\ 1063\AA\ outflow up to at least 17,000 km s$^{-1}$ can be probed.

\end{enumerate}

Among the resultant 1091 quasars, we search for relatively broad (FWHM$\gtrsim 500$ km s$^{-1}$) and deep troughs(minimum residual intensity less than 0.5) between the wavelength of the observed \civ\ and \siiv\ broad emission lines. This strategy avoids the narrower intervening systems and will find all \civ\ BALs and mini-BALs (see definitions in \S~4.1) with velocities between 0 and --30,000 \kms. We find such absorption systems in 191 objects. This fraction, 18\%, is close to the 16\% found by \citet{Dunn12} for similar-width troughs. 
We focus our detailed visual inspection on these 191 quasars, because the emergence of \siv\ absorption without a significant \civ\ trough 
is not expected \citep[see][]{Dunn12}, and the \civ\ region, by definition, has been the benchmark for characterizing absorption troughs in AGN outflows for
decades \citep[e.g.][]{Weymann81,Weymann91}.

\subsection{The VLT/X-shooter Sample}

Our SDSS/BOSS survey has the advantage of a large number of objects. However, the modest spectral resolution and S/N ratio of the 
SDSS/BOSS quasar data make the identification of the \siv\ and \siv* troughs challenging, especially for the narrower troughs. We therefore 
conducted a complementary survey using the X-shooter spectrograph on the VLT. To avoid selection biases that might give 
preference to a specific ratio of \siv*/\siv\ trough depths (and therefore $R$), we conducted a blind survey. We chose 13 SDSS objects at 
redshifts lower than 2.5 so that the SDSS data did not show the spectral region of the \siv\ and possibly \siv*\ troughs. Our two other selection criteria were (a) bright objects, to obtain a high S/N in a reasonable exposure time, and (b) deep \siiv\ outflow troughs, to increase the 
likelihood of detecting \siv\ and \siv*\ troughs.

Without prior knowledge of their spectral region, the \siv\ and \siv*\ troughs' depth ratio can fall into any of the categories described in 
sections 2.2.1, 2.2.2 and 2.2.3. For example, if a large majority of the outflows are situated at $R<1$ pc, photoionization models will 
require $\ne\gg\necrit$, and every \siv\ trough will be accompanied by an equal or deeper \siv* trough (see \S~2.2, and Fig.~\ref{fig:siv_sim}).
Seven out of the 13 objects in the X-shooter sample showed \siv\ outflow troughs (a total of eight outflows). 

Recent results from the full analysis of this VLT/X-shooter sample (Xu et al., 2018, ApJ submitted) show that the average value of $R$(\ne=\ncr)=250 pc for these eight outflows, where the individual $R$(\ne=\ncr) values span the range 80--600 pc.  This is in full agreement with the findings described in \S2.3.

\section{Identification of \siv\ absorption troughs}
\label{sec:Identification}

To identify absorption troughs associated with a quasar outflow, we begin with the \civ\ doublet $\lambda\lambda$1548.20, 1551.77. After identifying \civ\ trough(s) with velocities between 0 and --30,000 \kms\ in the rest frame of the quasar, we look for kinematically corresponding absorption features from the \siiv\ $\lambda\lambda$1394.76, 1403.77 doublet. We use the kinematic structure of the \siiv\ troughs 
as our guide for the existence of an \siv\ absorption trough. We do so for two reasons. First, experience shows that an \siv\ absorption trough is never detected if an \siiv\ trough is not detected; second, the 
\siiv\ trough profile shows good kinematic correspondence with the \siv\ trough profile \citep{Dunn12,Borguet12b,Borguet13,Chamberlain15a}. 

The existence of Ly$\alpha$ forest absorption troughs is 
the primary obstacle when trying to identify  \siv\ absorption troughs in the
spectra of the $z>2.6$ quasars. 
The intrinsic line widths of Ly$\alpha$ absorbers  are broadened
by the $\sim$160--180 km s$^{-1}$ spectral resolution of the SDSS and BOSS spectrographs (at the typical observed wavelength of \siv\ lines), as well as by random clumping of Ly$\alpha$ forest troughs.
To minimize the confusion due to the Ly$\alpha$ forest, we require that the \siiv\ 
troughs  have a velocity width of at least 500 \kms\ for the SDSS and BOSS data (we lessen this requirement for the higher-resolution VLT/X-shooter data). 

\subsection{Visual template fitting}

Often, the \siiv\ outflow troughs show a complex kinematic morphology. We identify a \siiv\ outflow component as a kinematic structure that has a local minimum of $I<0.7$, and whose width is larger than 500 \kms. Several of our surveyed quasars produce two and even three such \siiv\ outflow components.
In most cases, we create an AOD template for each individual component based on the 
\siiv~1402.77\AA\ trough. In the AOD case, the residual intensity is $I=e^{-\tau}$; therefore, our template is given by 

\begin{equation}
\tau(\lambda)_{1402.77}=-\ln[I(\lambda)_{1402.77}],
\label{eq:transform}
\end{equation}

where the subscript 1402.77 denotes that the absorption feature is identified as arising from the \siiv~1402.77\AA\ line.
In cases where the \siiv~1402.77\AA\ absorption is strongly blended with  the absorption of a different velocity component arising from 
the \siiv~1393.76\AA\ line, we use the  \siiv~1393.76\AA\ trough at the original velocity as a template.

We then make the assumption that the \siv\ trough will have the same kinematic profile as $\tau(\lambda)_{1402.77}$, scaled by a constant factor. This assumption works fairly well for cases where the saturation in the \siiv\ trough is less than a factor of two [i.e.,  $\tau$(real)$<$2$\tau$(AOD)].
In \S~\ref{sec:discussion} we elaborate on more saturated cases. 

To identify an \siv~1062.66\AA\ trough associated with  the outflow component identified by the \siiv\ absorption, we simply transform the template to the expected wavelength of the  \siv\ trough.

\begin{equation}
\lambda_{\rm \scriptstyle S\;{\scriptscriptstyle IV}\;expected\;trough}=\lambda_{\rm \scriptstyle Si\;{\scriptscriptstyle IV}\;trough}\frac{1062.66}{1402.77}.
\label{eq:transform2}
\end{equation}


A continuum-plus-emission-line model is a prerequisite for template-fitting as we need to work with residual intensities. We construct this model by visually depicting the overall continuum
with a power law and adding a Gaussian profile where an emission-line feature (the most relevant ones are associated with the \siiv$~\lambda\lambda$~1394,1403 and \ovi~$\lambda\lambda$~1032,1038 doublets) is present.
Although \ovi\ emission is assumed to be present in every case, if the spectrum is free of absorption lines, we adopt a conservative
strategy that when \ovi\ emission is completely unseen in the data, we do not artificially create it in this model. This avoids a significant arbitrariness and uncertainty in the derived emission model. 

When an emission model consisting of a power law and 
a number of Gaussians (at different wavelengths) fails to present a reasonable fit, we do a manual cubic spline fit instead. We then produce the residual intensity spectrum by dividing the fluxed spectrum by the emission model.

As described above, to account for 
the difference in optical depth between the \siiv\ and \siv\ troughs, we apply an SF to the $\tau(\lambda)_{1402.77}$, to best fit the  \siv\ absorption feature according to our visual inspection. 

Due to the frequent contamination from Ly$\alpha$ forest absorbers, sometimes
only a portion of the velocity profile of the \siiv\ template can be matched to the \siv\  trough. When this situation occurs, 
we limit our template to being situated predominantly on and above the actual data, because any portion of the template located below the actual 
data is nonphysical (with the uncertainty of spectroscopic measurement taken into account).
We then repeat the process for identifying a possible \siv*~1072.96\AA\ trough while allowing for a different SF than that used for the \siv\ trough. The error on the SF values is calculated using the average flux error per resolution element over the spectral region of the template (defined as $1\sigma$).  We calculate the SF associated with shifting the template by $\pm 1\sigma$ and report the difference between these values and the fitted SF as the errors.

In Table \ref{tab:SDSS}, we give details on the 25 quasars from our SDSS sample with detected \siv\ $\lambda1063$ absorption trough(s). Except for two, all the \civ\ troughs in these objects adhere to the width definition of a Broad Absorption Line (BAL); having a full width at 90\% residual intensity larger than 2000 \kms\ \citep[see][]{Weymann91}. The remaining two objects have \civ\ troughs with widths between 500 and 2000 \kms\ and therefore are classified as mini-BALs \citep[see][]{Hamann04}.
The two mini-BAL outflows are designated by "$m$" in front of their right ascension.

Table \ref{tab:VLT} gives similar parameters for quasars from our VLT/X-shooter sample with detected \siv\ $\lambda1063$ absorption trough(s). There are five BAL and two mini-BAL outflows in this sample (the latter are designated by "$m$" in front of their right ascension).

\begin{deluxetable}{r c c c r}[ht]

\setlength{\tabcolsep}{0.02in} 
\tablecaption{Our SDSS quasar sample with detected \siv\ $\lambda1063$ absorption trough(s).\label{tab:sdss1}}
\tablehead{
\colhead{RA, Dec} & \colhead{Plate-MJD-fiber} & \colhead{$z$} & \colhead{$r$}\\ 
\colhead{(1)} & \colhead{(2)} & \colhead{(3)} & \colhead{(4)} 
}

\startdata

07:57:15.34$\;+$21:33:33.8 & 4482-55617-760 & 2.965 & 18.17 \\
09:13:07.82$\;+$44:20:14.4 & 4687-56369-700 & 2.930 & 18.02 \\
09:37:10.77$\;+$25:57:20.0 & 5791-56255-652 & 2.711 & 18.41 \\
09:58:58.14$\;+$36:23:19.0 & 4637-55616-580 & 2.671 & 17.78 \\
11:31:32.86$\;+$49:28:14.0 & 6685-56412-853 & 2.625 & 18.14 \\
11:45:25.74$\;+$23:59:40.4 & 6422-56328-909 & 2.714 & 18.09 \\
$^*$11:45:48.38$\;+$39:37:46.7 & 4654-55659-856 & 3.108 & 17.94 \\
12:14:20.10$\;+$51:49:25.1 & 6682-56390-108 & 2.630 & 18.42 \\
$^m$12:20:17.06$\;+$45:49:41.2 & 6640-56385-266 & 3.275 & 18.18 \\
12:26:54.39$\;-$00:54:30.6 & 3792-55212-716 & 2.611 & 18.37 \\
12:37:54.83$\;+$08:41:06.8 & 5406-55955-204 & 2.859 & 17.47 \\
$^m$13:38:04.79$\;+$47:32:19.9 & 6748-56371-880 & 2.630 & 17.66 \\
$^*$13:47:22.83$\;+$46:54:28.5 & 6749-56370-348 & 2.915 & 17.98 \\
14:07:45.50$\;+$40:37:02.3 & 5169-56045-616 & 3.180 & 18.48 \\
14:22:40.79$\;+$21:26:44.4 & 5898-56045-232 & 3.023 & 18.47 \\
15:03:32.18$\;+$36:41:18.1 & 5168-56035-253 & 3.257 & 17.87 \\
$^*$15:09:23.38$\;+$24:32:43.3 & 6019-56074-110 & 3.045 & 17.48 \\		
15:43:54.87$\;+$49:27:21.4 & 6723-56428-436 & 2.625 & 17.67 \\	
15:50:47.65$\;+$58:07:34.8 & 6785-56487-216 & 3.070 & 18.33 \\	
16:42:19.88$\;+$44:51:24.0 & 6031-56091-588 & 2.893 & 17.86 \\
17:03:22.42$\;+$23:12:43.3 & 4177-55688-250 & 2.626 & 18.40 \\
22:30:06.55$\;+$21:59:08.5 & 6118-56189-832 & 2.870 & 18.30 \\
23:35:31.98$\;+$04:44:33.3 & 4283-55864-942 & 3.160 & 18.46 \\
23:41:50.01$\;+$14:49:06.1 & 6138-56598-406 & 3.165 & 18.43 \\
23:51:53.80$\;-$06:18:30.2 & 7146-56573-716 & 2.780 & 18.26
\enddata
\tablenotetext{*}{Objects already found by \citet{Dunn12}.}
\tablenotetext{$m$}{Objects classified as mini-BALs (see text).}
\tablecomments{
(1) Right ascension and declination (epoch 2000). 
(2) SDSS spectrum identifiers, including plug plate number, Julian Date, and fiber number.
(3) The quasar redshifts adopted here are those visually adjusted by \citet{Paris17} .
(4) The $r$-band magnitude obtained through PSF fitting.
}
\label{tab:SDSS}
\end{deluxetable}

\begin{deluxetable}{r c c c r}[ht]

\setlength{\tabcolsep}{0.02in} 
\tablecaption{Our VLT/X-Shooter quasar sample with detected \siv\ $\lambda1063$ absorption trough(s). \label{tab:vlt1}}
\tablehead{
\colhead{R.A., Decl.} & \colhead{Plate-MJD-Fiber} & \colhead{$z$} & \colhead{$r$}\\ 
\colhead{(1)} & \colhead{(2)} & \colhead{(3)} & \colhead{(4)} 
}

\startdata

00:46:13.54$\;+$01:04:25.8 & 4223-55451-666 & 2.149 & 18.04 \\
$^m$08:25:25.07$\;+$07:40:14.3 & 4866-55895-522 & 2.204 & 17.89 \\
08:31:26.16$\;+$03:54:08.1 & 4764-55646-244 & 2.076 & 18.27 \\
09:41:11.12$\;+$13:31:31.2 & 5318-55983-604 & 2.021 & 18.15 \\
$^m$11:11:10.15$\;+$14:37:57.1 & 5362-56017-170 & 2.138 & 18.03 \\
11:35:12.70$\;+$16:15:50.7 & 2503-53856-138 & 2.004 & 18.36 \\
15:12:49.29$\;+$11:19:29.4 & 1718-53850-564 & 2.109 & 17.65
\enddata
\tablenotetext{$m$}{Objects classified as mini-BALs (see text).}
\tablecomments{
(1) Right ascension and declination.
(2) SDSS spectrum identifiers, including plug plate number, Julian Date, and fiber number.
(3) The redshifts adopted here are those visually adjusted by \citet{Paris17}.
(4) The $r$-band magnitude obtained through PSF fitting.
}
\label{tab:VLT}
\end{deluxetable}

The manual template fits of all SDSS \siv\ absorption outflows are shown in Fig. \ref{fig:man_sdss1}, 
while the outflows in our VLT X-Shooter quasar sample are treated in the same way in Fig. \ref{fig:man_vlt}. The SF values denote the SFs (and their errors) by which we multiplied the $\tau(\lambda)_{1402.77}$ template to fit the \siv\ and \siv* troughs.

\begin{figure*}[]
\centering
\includegraphics[angle=90,scale=0.34,clip=true,trim=0mm 2mm 0mm 16mm]{sdss1a.ps}%
\includegraphics[angle=90,scale=0.34,clip=true,trim=0mm 0mm 0mm 16mm]{sdss1b.ps}\\
\caption{Visual template fitting. Each row shows a single outflow component, and the purple line shows the emission model. The right panel shows the \siiv\ outflow region, where the red line shows the absorption template for the \siiv~1402.77\AA\ line. In most cases, the blue line shows the fit of that template to the absorption feature of the same component 
caused by the stronger \siiv~1393.76\AA\ transition. The left panel shows the \siv\ and \siv* outflow region. In most cases, we transform the \siiv~1402.77\AA\
to the expected wavelength of the same component for the \siv\ and \siv* transitions. The blue trough shows the scaled \siiv~1402.77\AA\ template at the expected position for the \siv~1062.66\AA\ trough, and similarly, the red trough is for the expected position of the \siv*~1072.96\AA\ transition (see \S~4.1).
The SF that is applied to the AOD of the \siiv\ template is displayed for both the \siv\ and \siv*\ lines. In the text, we discuss the few cases where we use the absorption 
from the \siiv~1393.76\AA\ transition as a template, as well as the one case where we use the \aliii\ template (see inset figure for quasar SDSS J0937+2557). 
} 
\label{fig:man_sdss1}
\end{figure*}

\addtocounter{figure}{-1}

\begin{figure*}[]
\centering
\includegraphics[angle=90,scale=0.34,clip=true,trim=0mm 2mm 0mm 16mm]{sdss2a.ps}%
\includegraphics[angle=90,scale=0.34,clip=true,trim=0mm 0mm 0mm 16mm]{sdss2b.ps}\\
\caption{\tt Continued.} 
\end{figure*}

\addtocounter{figure}{-1}

\begin{figure*}[]
\centering
\includegraphics[angle=90,scale=0.34,clip=true,trim=0mm 2mm 0mm 16mm]{sdss3a.ps}%
\includegraphics[angle=90,scale=0.34,clip=true,trim=0mm 0mm 0mm 16mm]{sdss3b.ps}\\
\caption{\tt Continued.} 
\end{figure*}

\addtocounter{figure}{-1}

\begin{figure*}[]
\centering
\includegraphics[angle=90,scale=0.34,clip=true,trim=0mm 2mm 0mm 16mm]{sdss4a.ps}%
\includegraphics[angle=90,scale=0.34,clip=true,trim=0mm 0mm 0mm 16mm]{sdss4b.ps}\\
\caption{\tt Continued.} 
\end{figure*}

\begin{figure*}[ht!]
\centering
\includegraphics[angle=90,scale=0.34,clip=true,trim=130mm 2mm 0mm 0mm]{vlta.ps}%
\includegraphics[angle=90,scale=0.34,clip=true,trim=130mm 0mm 0mm 16mm]{vltb.ps}\\
\caption{Similar presentation as in Fig.\ \ref{fig:man_sdss1} but for our VLT X-Shooter quasar sample that shows \siv\ outflow troughs.} 
\label{fig:man_vlt}
\end{figure*}

Notes on two objects are as follows. \\
J0937+2557. This is a case of strong saturation in the \siiv\  troughs, for which we mention above that the template method is not applicable.
Therefore, we used the fact that \aliii\ troughs are seen in this outflow and are clearly not strongly saturated and constructed an \aliii\ template 
(see inset figure for the object in Fig.~\ref{fig:man_sdss1}), which fits the data very well. \\
J1503+3641. In the spectrum of this object, there is a clear damped \Lya\ system, so we normalized the spectrum accordingly (see Fig.~\ref{fig:man_sdss1})

\subsection{Quantifying the Goodness of Fit Using Monte Carlo Simulations}

\begin{figure}[t]
\centering
\includegraphics[angle=90,scale=0.22]{J1422+2126chiq-wavelength2.ps}
\caption{Example of our Monte Carlo simulations. Top panel: automated template fitting procedure shown for the expected positions of the \siv\ and \siv* troughs (blue and red, respectively) as well as for two random wavelength positions (green).
Bottom panel: $\chi^2_{mod}$ for the Monte Carlo fits at each random wavelength (purple symbols). The $\chi^2_{mod}$ values for the expected \siv\ and \siv* troughs are shown by the blue and red crosses, respectively, and by green crosses for the two Monte Carlo simulations show in the top panel. Yellow dots show the maximum optical depth of the fits at each wavelength.}
\label{fig:chiq-wavelength2}
\end{figure}

Figures \ref{fig:man_sdss1} and \ref{fig:man_vlt} show a compelling qualitative case for the identification of \siv\ troughs.
In all cases, the \siiv\ absorption template transformed to the expected wavelength of the \siv\ trough shows a good match with an observed absorption feature. The template can be fitted to the observed absorption feature using a single free parameter multiplying the \siiv\  optical depth template as a whole. In most cases, this procedure 
shows a compelling \siv\ trough at the expected position. 

To quantify the goodness of the
\siv\ template fits, we carry out Monte Carlo Simulations of automated template fitting. We select a wavelength range in the vicinity of the \siv/\siv*\ 
lines (typically 1038--1105\AA\ rest frame, avoiding the prominent absorption troughs arising from the \ovi~$\lambda\lambda$~1032, 1038 doublet) and transform the \siiv\ template  onto a random wavelength position within this range. We then progressively 
increase the depth of the template until any part of it reaches the actual spectral data. We consider  any portion of the template below the data
as unphysical, though we allow a $3\sigma$ tolerance level, where $\sigma$ is the uncertainty of the spectral flux density at each wavelength per pixel. 

In particular, we perform the following iterative fitting procedure.
\begin{itemize}
 \item We set the maximum optical depth (i.e. the deepest position) of the template to be 0.2 and determine if the template is below the actual 
data by $3\sigma$ or more within its wavelength range. 
 \item If this situation occurs, the iteration is stopped, and we consider the trough to be too shallow to be physical and abandon this fit.
 \item If the whole template is above the spectral data (minus 3$\sigma$), we successively increase the maximum optical depth of the template by 
0.01 each time. We iterate until the scaled template finally reaches the largest possible AOD before any portion  dips $3\sigma$ below the data, which we take as the final fitting result.  We then calculate a goodness-of-fit  figure of merit, which is a modified reduced $\chi^2$ ($\chi^2_{mod}$; see below) for this final fit to the data.
\end{itemize}

We perform the above procedure at 1000 random wavelengths in the chosen 1038--1105\AA\ rest-frame range. 
We then use the same procedure for the template that is transformed to the exact position expected for an \siv\ 1062.66\AA\ trough (using equation (\ref{eq:transform2})) and calculate the  $\chi^2_{mod}$ of this ``expected trough fit'' as well. We do the same for the template that is transformed to the exact position expected for an \siv*~1072.96\AA\ trough.

\subsubsection{Goodness-of-fit figure of merit}

The identification of a of an \siv\ trough amidst the \Lya\ forest absorption rests on three considerations. \\
1)	A significant absorption feature has to coincide with the exact spectral position predicted by the \siiv\ template (using equation (\ref{eq:transform2})). \\
2)	Not only does an absorption feature need to exist at that predicted wavelength position, but its shape has to match a significant portion of the scaled template. \\
3) Deeper absorption features adhering to points 1 and 2 are better candidates for an \siv\ trough amongst the myriad \Lya\ forest absorption features.

Using the standard $\chi^2=(model-I)^2/\sigma^2$ (where $I$ is the residual intensity at the particular wavelength, model is the value of predicted 
$I$ using the scaled template at the same wavelength, and $\sigma$ is the error on the measurement) yields lower $\chi^2$ values for shallow features when systematic effects dominate the fit. For example, if $I>0.8$, then using our fitting procedure $(model-I)^2<0.04$ whereas for a deep feature with  $I>0.2$,  $(model-I)^2<0.64$. Thus, inherently, shallow features will tend to have lower values of  $\chi^2$. 

We therefore introduce a modified $\chi^2$
($\chi^2_{mod}$) that compensates for this deep trough effect by normalizing each point by  the depth of the data.
\begin{equation}
 \chi^2_{mod}\equiv\frac{1}{n-1}\sum_j\left[\frac{model(j)-I(j)}{\sigma(j)}\times\frac{1}{1-I(j)}\right]^2,
\end{equation}
where the sum $J$ is taken over all the the points of the template and $n$ is the number of degrees of freedom.
It is important to note that the use of $\chi^2_{mod}$ as a goodness-of-fit measure does not give preference for \siv\ absorption features.  Instead, it gives preference for deeper absorption features of all kinds, as is desirable when trying to identify distinct absorption features in data dominated by \Lya\ forest absorption features (see point 3 above)


The distribution of $\chi^2_{mod}$  from the 1000 random wavelengths is compared to the reduced $\chi^2_{mod}$
value of the expected trough fits.  Cases where less than 10\% of the random fits have smaller $\chi^2_{mod}$ than the expected trough fits
are considered to be good fits for a \siv\ outflow trough.

 Many times we find good fits for both  \siv\ and  \siv* outflow troughs.  
The chance probability of a random 
combined \siv\ and  \siv* outflow trough fit is much lower than that for an individual trough fit.
Many such cases can be seen in figures \ref{fig:man_sdss1} and \ref{fig:man_vlt}, which significantly strengthen the viability of the identification of theses troughs amidst the \Lya\ forest.




In figure \ref{fig:chiq-wavelength2} we show an example of these simulations, 
and in Tables \ref{table_sdss_results} and  \ref{table_vlt_results} we give this representative percentage  for each outflow component,
which gives a quantitative measure for the goodness of the template fit. 


\section{Results}
\label{sec:results}

\begin{deluxetable*}{c r r r c c c c l}[htb!]
\tablewidth{\textwidth}
\tabletypesize{\small}
\setlength{\tabcolsep}{0.02in}
\tablecaption{Derived properties of \siv\ and \siv*\ Troughs in the SDSS quasar sample. \label{tab:sdss2}}
\tablehead{
\colhead{Object} & \colhead{$v_{\rm ctr}$} & \colhead{$\Delta v$} & \colhead{$f(\chi^2)$} 
& \colhead{$1.2<\frac{N(\rm S\;{\scriptscriptstyle IV^*})}{N(\rm S\; \scriptscriptstyle IV)}$}
& \colhead{$0.8<\frac{N(\rm S\;{\scriptscriptstyle IV^*})}{N(\rm S\; \scriptscriptstyle IV)}$$<$1.2}
& \colhead{0.5$\leqslant$$\frac{N(\rm S\;{\scriptscriptstyle IV^*})}{N(\rm S\; \scriptscriptstyle IV)}$$<$0.8}
& \colhead{$\frac{N(\rm S\;{\scriptscriptstyle IV^*})}{N(\rm S\; \scriptscriptstyle IV)}$$<$0.5} \\
\\ [-2mm]
\colhead{(1)} & \colhead{(2)} & \colhead{(3)} & \colhead{(4)} & \colhead{(5)} & \colhead{(6)} & \colhead{(7)} & \colhead{(8)} 
}

\startdata
J0757$+$2133   &  $-$1950 & 1400 & 9.4\% &             & 1.00$\pm$0.15 &         &            \\ 
J0913$+$4420   &  $-$3440 & 1600 &  0.3\% &             & 0.90$\pm$0.09 &         &            \\ 
J0937$+$2557   &  $-$3160 & 2100 & 3.7\% &             &          & 0.53$\pm$0.08 &            \\ 
J0958$+$3623   &  $-$2950 & 1100 &  2.8\% &             &      & 0.50$\pm$0.09 &            \\ 
\ \ \ J1131$+$4928 A &  $-$4070 &  700 &  0.4\% &             &      &         & 0.42$\pm$0.07 \\ 
\ \ \ J1131$+$4928 B &  $-$830  &  900 &  0.2\% &             &      &         & 0.08$\pm$0.07 \\ 
J1145$+$2359   & $-$13340 & 3000 & $(a)$3.7\% &             & 0.80$\pm$0.12 &         &            \\ 
J1145$+$3937   &  $-$4570 & 1800 & 4.3\% &             & 1.00$\pm$0.12 &         &            \\ 
\ \ \ J1214$+$5149 A &  $-$6900 & 1200 &  0.3\% &             & 0.82$\pm$0.28 &          &            \\ 
\ \ \ J1214$+$5149 B &  $-$2900 & 1900 &  0.9\% &             &      &         & 0.33$\pm$0.12 \\ 
J1220$+$4549   &   $-$220 & 1100 &   0.5\% &             &      &         & 0.07$\pm$0.04 \\ 
J1226$-$0054   &  $-$2570 & 1800 &  3.0\% &             & 1.20$\pm$0.33 &         &            \\ 
\ \ \ J1237$+$0841 A &  $-$3700 &  900 &  6.8\% &             & 0.83$\pm$0.08 &         &            \\ 
\ \ \ J1237$+$0841 B &  $-$1060 &  600 & 8.4\% &             &      &  &    0.44$\pm$0.06        \\ 
J1338$+$4732   &  $-$1860 & 1100 & 3.3\% &             &      &         & 0.31$\pm$0.17 \\ 
\ \ \ J1347$+$4654 A &  $-$5610 & 1300 &  0.2\% &      $(b)$3.20$\pm$0.45       &  &         &            \\ 
\ \ \ J1347$+$4654 B &  $-$1670 & 2300 &  0.8\% &             & 1.17$\pm$0.11 &         &            \\ 
\ \ \ J1407$+$4037$(c)$   & $-$11710 & 3400 & 8.6\% &   1.50$\pm$0.15  &       &         &          \\ 
J1422$+$2126   &  $-$3890 & 1100 & 3.8\% &             &      &         & 0.17$\pm$0.04 \\                                      
J1503$+$3641   &  $-$4550 & 1800 & 7.3\% &             &      &         & 0.40$\pm$0.10 \\ 
\ \ \ J1509$+$2432 A &  $-$2770 & 2400 &   3.2\% &   1.41$\pm$0.09 &      &         &            \\ 
\ \ \ J1509$+$2432 B &  $-$1180 &  700 &   8.0\% &             &      & 0.58$\pm$0.04 &            \\ 
\ \ \ J1509$+$2432 C &  $-$418  &  700 &  3.6\% &             &      &         & 0.15$\pm$0.03 \\ 
\ \ \ J1543$+$4927 A &  $-$9460 & 1400 &  2.9\% &             &      &         & 0.35$\pm$0.19 \\ 
\ \ \ J1543$+$4927 B &  $-$6010 &  700 &   3.4\% &             & 0.82$\pm$0.25 &         &            \\ 
\ \ \ J1543$+$4927 C &  $-$4210 & 1300 &  8.3\% &             &      &         & 0.12$\pm$0.24 \\ 
J1550$+$5807   &  $-$3230 &  800 &  5.2\% &             &0.80$\pm$0.16 &         &          \\ 
J1642$+$4451   &  $-$6820 &  900 &  0.3\% &             &      &         & 0.33$\pm$0.15 \\ 
\ \ \ J1703$+$2312 A &  $-$5260 & 1500 & 2.2\% &             & 0.83$\pm$0.13 &         &          \\ 
\ \ \ J1703$+$2312 B &  $-$1530 &  700 &  1.6\% &             &      &         & 0.21$\pm$0.04 \\ 
J2230$+$2159   &  $-$8980 & 1100 &  5.7\% &             & 1.00$\pm$0.22 &         &            \\ 
J2335$+$0444   &  $-$3510 &  900 &  6.5\% &    1.71$\pm$0.50 &      &         &            \\ 
J2341$+$1449   &  $-$6220 &  500 & 3.9\% &             &      &         & 0.06$\pm$0.02 \\ 
J2351$-$0618   &  $-$1850 &  700 & 4.2\% &             &      & 0.50$\pm$0.12 &            \\ 
{\bf Total population} &    &         &       &    {\bf 4}         &    {\bf 12}       &     {\bf 4}   &   {\bf 14} \\
\vspace{-2.2mm}
\enddata

\tablecomments{
(1) Abbreviated object name and outflow component.
(2) Relativistic velocity (in km s$^{-1}$) of the \siv\ outflow in the quasar's rest frame, measured at the deepest part of the absorption trough.
(3) Full width at 90\% of the \siv\ trough, measured from the  line profile template constructed from the \siiv\ trough.
(4) Fraction of Monte Carlo template fits giving a $\chi^2_{mod}$ smaller than the one obtained at the expected wavelength of the \siv\ 1062.66\AA\  trough (obtained by 
using the same automated fitting procedure as our Monte Carlo simulation).
(5)--(8) Value of $N(\rm S\;{\scriptstyle IV}^*)_{\rm AOD}/N(\rm S\;{\scriptstyle IV})_{\rm AOD}$, split among the four categories discussed in \S~5.1. These values are calculated from the SFs of both troughs in our manual template fits shown in  Figure \ref{fig:man_sdss1}. For $(a)$, $(b)$ and $(c)$ - see \S~5.1.1
}
\label{table_sdss_results}
\end{deluxetable*}

\begin{deluxetable*}{l r r r c c c c}[htb!]
\tablewidth{\textwidth}
\tabletypesize{\small}
\setlength{\tabcolsep}{0.02in}
\tablecaption{Derived properties of \siv\ and \siv*\ troughs in the VLT/XShooter quasar sample. \label{tab:vlt2}}
\tablehead{
\colhead{Object} & \colhead{$v_{\rm ctr}$} & \colhead{$\Delta v$} & \colhead{$f(\chi^2)$} 
& \colhead{$1.2<\frac{N(\rm S\;{\scriptscriptstyle IV^*})}{N(\rm S\; \scriptscriptstyle IV)}$}
& \colhead{$0.8<\frac{N(\rm S\;{\scriptscriptstyle IV^*})}{N(\rm S\; \scriptscriptstyle IV)}$$<$1.2}
& \colhead{0.5$\leqslant$$\frac{N(\rm S\;{\scriptscriptstyle IV^*})}{N(\rm S\; \scriptscriptstyle IV)}$$<$0.8}
& \colhead{$\frac{N(\rm S\;{\scriptscriptstyle IV^*})}{N(\rm S\; \scriptscriptstyle IV)}$$<$0.5} \\
\\ [-2mm]
\colhead{(1)} & \colhead{(2)} & \colhead{(3)} & \colhead{(4)} & \colhead{(5)} & \colhead{(6)} & \colhead{(7)} & \colhead{(8)} 
}

\startdata
J0046$+$0104   &  $-$1746 &  606 & 1.4\% &          &           &          & 0.36$\pm$0.06 \\ 
J0825$+$0740   &   $+$406 &  200 &  0.4\% &          &           & 0.60$\pm$0.18 &          \\ 
J0831$+$0354   & $-$10800 & 1350 & 0.3\% &          &           & 0.72$\pm$0.07 &          \\ 
J0941$+$1331   &  $-$4874 &  329 & 6.8\% &          &           &          & 0.29$\pm$0.12 \\ 
J1111$+$1437   &  $-$1846 &  666 &  1.6\% &          &           &          & 0.27$\pm$0.04 \\ 
J1135$+$1615   &  $-$6754 & 1332 &  4.4\% & 1.33$\pm$0.54 &          &        &          \\ 
J1512$+$1119 A &  $-$1876 &  335 &  6.2\% &          & 1.00$\pm$0.05 &        &          \\ 
J1512$+$1119 B &  $-$1053 &  322 & 1.8\% &          &          &         & 0.08$\pm$0.09 \\ 
{\bf Total population} &    &         &       &    {\bf 1}       &   {\bf 1}     &    {\bf 2}    &   {\bf 4} \\
\vspace{-2.2mm}
\enddata

\tablecomments{
Same as Table \ref{table_sdss_results} for the VLT/XShooter sample.
}
\label{table_vlt_results}
\end{deluxetable*}


\subsection{SDSS Sample}

As we noted in \S~3.1, from 1091 SDSS and BOSS quasars in our sample, 191 quasars show \civ\ outflow absorption troughs with velocity widths larger than 500 \kms. Of these 191 quasars, we found 25 that show robust detection of corresponding \siv\ outflow troughs, or 13\%. This fraction is a lower limit, since (a) in many cases 
the data allow for the existence of an  \siv\  trough, but a kinematic correspondence with the \siiv\ trough does not exist due to strong blending with Ly$\alpha$ forest absorption troughs; (b) due to the thick Ly$\alpha$ forest and limited S/N and spectral resolution,  we do not identify \siv\  troughs with optical depths less than 
0.3, therefore shallow \siv\ troughs are not identified.
Several of the \siv\ quasars show two or three distinct outflow components, so in total, our survey yielded
34 \siv\ outflow components. 

The analysis results for these components are shown in Table \ref{table_sdss_results}. 
In section 2 we demonstrated the following points. \\ 
1) For luminous quasars, outflows with $n_e<\ncr$ (where \ncr\ is the critical density for the excited and resonance energy levels of \siv) are situated at distances larger than 100 pc from the central source (see section 2.3).
\\
2) When $N(\rm S\;{\scriptstyle IV}^*)_{\rm AOD}/N(\rm S\;{\scriptstyle IV})_{\rm AOD}<1$, that is when the \siv*\ trough is shallower than the \siv\ trough,  
$n_e<\ncr$ (see \S~2.2). 
(Hereafter, we use $N$(\siv*)/$N$(\siv) to donate $N(\rm S\;{\scriptstyle IV}^*)_{\rm AOD}/N(\rm S\;{\scriptstyle IV})_{\rm AOD}$, including in Tables 3 and 4. )\\
Since $N$(\siv*)/$N$(\siv) equals the ratio of the \siv\ to \siv*\ template SFs 
(see \S~3), we can calculate the values for the former using the SF given in figure~3. In Table 4, we separate the values of $N$(\siv*)/$N$(\siv) into four physically motivated columns: \\ 
Column 5 gives $N$(\siv*)/$N$(\siv)$>1.2$, for which $n_e>\ncr$ and therefore $R<$~100 pc (corresponds to the last row in Table 1)\\
Column 6 gives $N$(\siv*)/$N$(\siv)$\sim1$, for which $R$ can be smaller 100 pc.\ (the full constraints are discussed in \S2.2.3 and summarized in the third row of Table 1). Taking into consideration the errors on $N$(\siv*)/$N$(\siv)$\sim1$, in this case, we grouped together in this case all of the outflows with $0.8<N$(\siv*)/$N$(\siv)$<1.2$. \\
Column 7 and 8 give $N$(\siv*)/$N$(\siv)$<0.8$, for which we are certain that $n_e<\ncr$ and therefore $R>$~100 pc (corresponds to the first two rows in Table 1). We further split this category into two columns, where the outflows in column 8 have $N$(\siv*)/$N$(\siv)$<0.5$, where no reasonable error can bring the actual ratio to be consistent with $n_e\geq\ncr$.

We find that 53\% (18 out of 34) of the outflows have $N$(\siv*)/$N$(\siv)$<0.8$, which situates them at distances of more than 100 pc\ from the central source (see \S~2.3). In 14 outflow components, this assertion is more robust, as they have $N$(\siv*)/$N$(\siv)$<0.5$.

 Some of our \siv\ troughs might be false-positive identifications due to the \Lya\ forest contamination. To address the effect of this issue on our results, we use the Monte Carlo simulation statistical figure of merit (see Table \ref{table_sdss_results}). On average, only 3.7\% of our Monte Carlo simulations give better template fits than the fit at the exact \siv\ expected wavelength. Therefore, we expect only about one of the 34 identified \siv\ troughs to be a false detection, on average. 

To check the empirical situation, we have done the reverse experiment as well. We attempted to identify \siv* troughs in the population of objects where we did not identify \siv\ troughs, that is objects from the original 191 \civ\ detections (see \S~3.1) that do not appear in Table 2 and figure 3 (166 quasars in total). We use the same identification procedure that was used in \S~4.1 to identify \siv* troughs. First, we identify an \siiv\ outflow component as a kinematic structure that has a local minimum of $I<0.7$, and whose width is larger than 500 \kms; 32 out of the 166 objects showed such components in their spectra. Several of these 32 quasars produce two and even three such \siiv\ outflow components, for a total of 43 components. Only one of these 43 components showed a good match for an \siv* trough:\\
SDSS J1208+0020, where a good fit can be found for an \siv* trough with an SF=1 ($\Delta v=700$ \kms\ and maximum $\tau_{AOD}$=0.5). The data do not allow any \siv\ trough.  \\
Finding only one false-positive detection testifies for the accuracy of our \siv\ trough identification method. Furthermore, finding one false identification in 43 components is in good agreement with the Monte Carlo-based 3.7\% chance of false identification discussed in the previous paragraph.

In figure \ref{fig:dv-r-v-r}  we show the distribution of $N$(\siv*)/$N$(\siv) vs.\ the outflow velocity and velocity width. We find no correlation of $N$(\siv*)/$N$(\siv) with the outflow's velocity.
We do find some correlation between $N$(\siv*)/$N$(\siv) and the velocity width of the outflows, as all four outflows with $\Delta v>2000$ \kms\ are consistent with $N$(\siv*)/$N$(\siv)$=1$

\subsubsection{Notes for individual outflows} 

\noindent $(a) - $ J1145$+$2359 is the one case where we used the percentage fit of the \siv*\ trough. This was done because the \siv\ and \siv*\ troughs are so wide that they self-blend, leaving the low velocity of the \siv*\ trough as the only clearly identifiable feature while using the \siiv\ template. 

\noindent $(b) - $ For the outflow J1347$+$4654A, the ratio $N$(\siv*)/$N$(\siv)$=3.2$ is unphysical (as the maximum can only be 2).  This high value arises because what is fitted as the \siv* trough for component A is mostly due to the   \siv\ trough for component B; see Fig.~3.

\noindent $(c) - $ J1407+4037 has the widest BAL in the \siv\ sample, with a \civ\ balnicity of 17,000 km~s$^{-1}$, and it has a deep and very wide blended \siv/\siv* BAL (see Fig. 3). The \siiv\ trough of this wide outflow is the only one in the \siv\ sample that does not show distinct components. We did not want to exclude this object from the analysis, as it will bias the sample against the widest BALs. Therefore, we fit the high-velocity portion of the \siiv\ trough (which is the least-blended portion of the trough) and use it as a template for the \siv\ region. The width of the template was chosen to be 2900 km s$^{-1}$ as that is the velocity separation between the \siv\ and \siv* transitions.

\begin{figure}[t]
\centering
\includegraphics[angle=90,scale=0.13]{dv-r-v-r.ps}
\caption{Possible correlations for the SDSS sample. Top panel: Velocity width of the \siv\ outflow components as a function of $N$(\siv*)/$N$(\siv*). Bottom panel: velocity centroid of the \siv\ outflow components as a function of $N$(\siv)/$N$(\siv*). All outflows in Table \ref{table_sdss_results} are shown in both panels.} 
\label{fig:dv-r-v-r}
\end{figure}

\subsection{VLT/X-shooter Sample}

The VLT/X-shooter data has three times the spectral resolution and two to three times the S/N of the SDSS data.
Therefore, we look for \siv\ absorption troughs that accompany an \siiv\ trough wider than 150 \kms. From the right identified \siv\ outflow troughs, four are narrower than 500 \kms. 
Seven of these outflows are shown in figure \ref{fig:man_vlt} and another outflow (in SDSS J0831+0354) is analyzed and discussed intensively in \citet{Chamberlain15a}

We find that six (75\%) of the right outflows have $N$(\siv*)/$N$(\siv)$<1$, which situates them at distances of more than 100 pc\ from the central source (see \S~2.3). While the sample is small, it suggests that better data resolve some of the ambiguity due to \Lya\ forest contamination and reveals a higher percentage of outflows with $R>$100~pc.

\subsection{Very Large Scale Outflows}

Four of the 34 SDSS/BOSS and one of the right X-shooter outflows have no detected \siv* trough;
thus, we can put a conservative upper limit of $N$(\siv*)/$N$(\siv)$<0.1$. These outflows tend to have smaller  
values of $N$(\siv)$_{\mathrm{AOD}}$, resulting in lower values of \uh\ and therefore larger distances for a given   $N$(\siv*)/$N$(\siv). As a result, the $N$(\siv*)/$N$(\siv)$<0.1$ outflows have $R>1000$ pc. For example, a detailed photoionization solution for component B in SDSS 1512+1119 was published by \cite{Borguet13}, where an upper limit of $N$(\siv*)/$N$(\siv)$<0.1$ was converted to a lower limit of $R>3300$~pc.

\subsection{What Is the R Value for Outflows with N(\siv*)/N(\siv)$\geq 1$?}

When $N$(\siv*)/$N$(\siv)$\geq 1$, we only have an upper limit on $R$. Therefore, in principle, $R$ can be as small as the radius of the accretion disk around the supermassive black hole ($R\leq0.01$ pc). However, there are two empirical observations that suggest larger $R$ values in this case.
First, there is one case where $N$(\siv*)/$N$(\siv)$\geq1$ is found, but there exist other high-quality excited-state trough diagnostics that are sensitive to \ne$>$\ncr(\siv). In that object (SDSS J1512+1119), VLT/X-shooter data show identical \siv~1062.66\AA\ and \siv*~1072.96\AA\ troughs 
(see \citealt{Borguet12b}, Fig. 4). In the same, data we also detect \ciii* troughs that allow us to constrain $10^{4.8}\leq\ne\leq10^{8.1}$ (cm$^{-3}$)
(see \citealt{Borguet12b}, Fig. 5), which leads to $10$ pc $<R<300$ pc  \citep{Borguet13}. This excludes wind accelerated from the  accretion disk as the origin of that outflow.  

A more general argument comes from the similarity of all \siv\ outflows. The observed  velocity and velocity width of the troughs are quite similar for objects with $N$(\siv*)/$N$(\siv)$<1$ ($R>100$~pc) and objects with $N$(\siv*)/$N$(\siv)$\geq1$. This is evident in figure \ref{fig:dv-r-v-r} (where errors in determining $v$ and $\Delta v$ are less than 100 \kms).
Such characteristics are easier to explain if more of the  $N$(\siv*)/$N$(\siv)$\geq1$ outflows have $100$ pc $>R>10$~pc than $1$ pc $>R>0.01$~pc, as outflows with the latter range of $R$ require much more fine-tuning to look like their relatives at $R>100$~pc. This is because photoionization arguments require $\ne\propto R^{-2}$ and most acceleration mechanisms tend to yield a higher velocity for outflows originating at smaller $R$ (for example, $v\propto R^{-1/2}$ is typical for radiative acceleration scenarios; (e.g., \citealt{Arav94}).

\section{Extrapolating the \siv\ results to the general population of HiBAL outflows}
\label{sec:extrapolation}


As described above, we found that half of the \siv\ outflows found in our survey are situated at distances larger than 100 pc from the central source. Here we detail the steps and assumptions that are needed to extend these results to the general population of HiBAL outflows.


\subsection{Extrapolation to all \civ\ BALs in our sample}
In the parent sample of 1091 quasars, we identify 130 \civ\ absorption components that qualify for the width definition of a BAL \citep[see][]{Weymann91}. 
We note that our BAL detection fraction (130/1091=12\%) is similar to the detection rate of \civ\ BALs in optical surveys: e.g., 15$\pm3$\% found for the large bright quasar survey between redshifts 1.5 and 3
\citep{Hewett03} and 10.4$\pm0.2$\% in the SDSS DR3 catalog for the redshift
range 1.7 -- 4.38,  
\citep{Trump06}.
Therefore, our selection criteria of \civ\ trough with minimum
residual intensity (\textit{I}) less than 0.5 does not miss a significant population of BALs. 
However, we note that our results are not applicable to \civ\ BAL systems with minimal $I>0.5$.

In 32 of these 130 \civ\ BALs, which represents 25\% of the total BAL outflows, we identified 
\siv\ absorption features (see Tables 2 and 4).
The first issue is: can the \siv\ results be extrapolated to the 75\% of the \civ\ BAL outflows in our sample where there is no clear detection of corresponding \siv\ absorption?

As shown by \citet{Dunn12}, in photoionization equilibrium, 
the ratio of \siv\ to \civ\  ionic column densities is rather constant (to within a factor of 3) as a function of the photoionization parameter ($U_H$) in the range $-4<\log(U_H)<-1$. This demonstrates that \civ\ and \siv\ are formed in the same physical region of the outflow. However, for solar metallicity, 
the expected optical depth ratio 
$\tau$(\siv~$\lambda1063$)/$\tau$(\civ~$\lambda1548$)$\sim$0.01    for the same span of $U_H$ values.
This small ratio is due the lower abundance of sulfur compared to carbon (roughly 1/16 for solar abundances), a ratio of 1/4 in oscillator strengths, and a ratio of 2/3 in wavelength ratio \citep[see Eq. (2) in][]{Dunn12}.


The small expected $\tau$ ratio explains the absence of \siv\ absorption associated with 75\% of the \civ\ BALs in our sample. For a given $U_H$ value, there is a large range of total hydrogen column density ($N_H$)
that produces a BAL with $\tau$(\civ) between 0.5 and 30, which is easily detectable.
However, since for the same $N_H$ $\tau$(\siv)/$\tau$(\civ)$\sim$0.01  (see above paragraph), $\tau$(\siv) is between roughly 0.005 and 0.3, 
which is undetectable amidst the \Lya\ forest absorption features, even at the high-end value. Only outflows with larger $N_H$ would produce detectable \siv\ absorption associated with the same BAL.

Therefore, to extend the \siv\ results to the 75\% of the \civ\ BAL outflows in our sample where there is no clear detection of corresponding \siv\ absorption, we need two assumptions.\\
1) The difference in \civ\ vs. \siv\ absorption detection is mainly due to the latter needing a larger $N_H$ to be detected. 


We note that the larger $N_H$ needed for the detection of \siv\ absorption also explains the higher fraction of observed \siiv\ and \aliii\ in these outflows.

2) Outflows with lower $N_H$ (where it is difficult to detect \siv\ absorption troughs) are not preferentially found at smaller distances than outflows with higher $N_H$. 

To address this issue, we examine the relationship between $R$ and $N_H$ for outflows where $R$ was determined using doubly and triply ionized species (see discussion in the Introduction). 
Table \ref{tab:hiionR} shows the published parameters of these outflows. The trend is for lower $N_H$
outflows to be found at larger $R$ than outflows with higher $N_H$.  We also note that with high-quality 
data, it is possible to detect \siv\ outflows with lower $N_H$.  The one example we have is 
SDSS J1512+1119B (based on VLT/X-shooter data), which has the second-lowest  $N_H$ in Table \ref{tab:hiionR}
and the second-largest $R$.  We conclude that the data at hand do not support the possibility that outflows with lower $N_H$  are found preferentially at smaller $R$ than outflows with higher $N_H$.

\begin{deluxetable}{l c c r r r}[ht]

\setlength{\tabcolsep}{0.02in} 
\tablecaption{Outflows with Published Distances Based on High-ionization Dignostics. \label{tab:highionR}}
\tablehead{
\colhead{Object} & \colhead{Distance} & \colhead{$\log(N_H)$} & \colhead{$R$} & \colhead{$Ref$}\\ 
\colhead{(1)} & \colhead{Diagnostic} & \colhead{(cm$^{-2}$)} & \colhead{(pc)} & \colhead{}
}

\startdata

SDSS J0831+0354 & \siv\ & 22.5 & 110 & 1 \\
SDSS J1106+1939 & \siv\ & 22.1 & 320 & 2 \\
SDSS J1512+1119A & \siv\ and \ciii\ & 21.9 & 10-300 & 2\\
SDSS J1111+1437 & \siv\ & 21.5 & 880 & 3 \\
HE0238--1904 & \oiv\ & 20.7 & 1700 & 4 \\
SDSS J1206+1052 & \niii\ and \siii\ & 20.5 & 840 & 5 \\
SDSS J1512+1119B & \siv\ & 20.1 & $>$3000 & 2\\
FBQS J0209--0438 & \oiv\ & 20.0 & 4000 & 6
\enddata
\tablecomments{References 
1: \citet{Chamberlain15a};
2: \citet{Borguet13};
3:  Xu et al. (2018);
4: \citet{Arav13};
5:  \citet{Chamberlain15b};
6: \citet{Finn14}.
}
\label{tab:hiionR}
\end{deluxetable}

\subsection{Extending the Results to Fainter and Lower-Redshift Objects }

Due to instrument and S/N considerations, our parent sample of 1091 objects is limited to a redshift above 2.6 and an {\it r} magnitude larger than 18.5.  Extrapolating the \siv\ results to lower redshift and fainter objects implicitly assumes that the physical nature of the outflows does not depend strongly on the redshift of brightness of the observed objects.
To date, we are unaware of published results suggesting that the physical nature of BAL outflows depends on the brightness or redshift of the observed quasar.  
 
\section{discussion}
\label{sec:discussion}

\subsection{Comparison of Distance Determination Methods for AGN Outflows}

In this paper, we concentrated on the excited-state trough method for finding the distance of the outflow from the central source ($R$). 
In the literature, we find four other methods for deducing $R$. Here we describe these methods and compare their advantages and disadvantages 
to those of the excited-state trough method.

(1) The most robust method for $R$ determination uses spatially resolved spectroscopy (usually integral field unit spectroscopy (IFU)) 
to directly measure the size of the outflow across the field of view. This method is used extensively in the study of Seyfert outflows 
\citep[e.g.][]{Barbosa09,Riffel11}, 
as well as for outflows in luminous quasars \citep[e.g.,][]{Liu13a,Liu13b,Liu14,Liu15,CanoDiaz12,Harrison12,Harrison14,McElroy15}. A typical spatial 
resolution of $\geqslant$0.5\arcsec\ limits this method to large-scale outflows: a few tens of pc in nearby AGNs and several thousand pc for luminous quasars at redshifts larger than $\sim0.5$. Such outflows are readily observed with these IFU observations. For luminous quasars, typical outflow values are   $R~15,000$~pc and $v\ltorder1000$~km~s$^{-1}$. The drawback of this method is its current inability to probe luminous quasar outflows 
on scales of less than several thousand pc for luminous quasars at redshifts larger than 0.5, even using adaptive optics. Observing larger-scale outflows using  IFU data does not exclude the possibility of outflows on much smaller $R$ values.

Two other methods use trough variability. The idea is to translate the time-scale of variability seen in the outflow troughs 
\citep[e.g.][]{Barlow92,FilizAk13,Grier16,Rogerson16,Matthews16} into constraints on $R$. These efforts are dependent on the mechanism assumed to cause the trough's variability. There are two main mechanisms invoked to explain trough variability. 

(2) Changes in the ionizing flux incident on the outflow alter the ionization equilibrium, which increases or decreases the fractional abundance of a given ion. Such an occurrence does not necessitate a change in the total column density of the absorber. Since photoionization changes are a time-dependent process, the time-scale for variability  (or lack thereof) can be used to extract constraints on the electron number density \citep[\ne; e.g.,][]{Krolik01,Arav12,Arav15}. These \ne\ constraints, combined with a photoionization solution for the outflow, yield constraints on $R$ (similar to the method described in  \S~2.3). Uncertainties on the ionizing flux light curve, as well as decreasing variability on smaller timescales, causes this method to be less reliable than the excited-state trough method.

(3) Trough changes are attributed to material crossing the line of sight. With an estimate of the emission region's size and assuming a Keplerian-dominated tangential motion for the absorbing material, it is possible to constrain $R$ using the variability time-scale of the trough \citep[e.g.,][]{Moe09, 
Capellupo11,Capellupo12}. In general, such an occurrence requires a change in the total column density of the absorber. The assumption of  Keplerian-dominated tangential motion and the lack of a self-consistent model for the changes in the troughs' depth cause this method to be less reliable than the excited-state trough method (see \S~2).

Unfortunately, the two mechanisms invoked to explain trough variability provide very different $R$ estimates; therefore, unless we can ascertain the underlying mechanism for trough variability, it is impossible to deduce robust constraints on $R$ from studying trough variability. The best-studied case of AGN trough variability is from two monitoring campaigns of NGC~5548, where many simultaneous determinations of UV trough shapes, the UV broadband flux curve, and X-ray flux are available \citep[see][]{Kaastra14,Peterson14}.  For this case, the   
changes in the ionizing flux model yield $R$ estimates fully consistent with the excited-state trough method 
\citep{Arav15}, while the  material crossing the line-of-sight model is inconsistent with the $R$ derived from the excited-state trough method.

For BAL outflows, recent analysis suggests that variation of the ionizing continuum is the main driver of most observed BAL variability \citep{He17}. Two published analyses of individual objects that attempt to decide the variability mechanism support a different interpretation in each case: \citet{Capellupo14} argued for the material crossing the line-of-sight explanation, while   \citet{Sternd17} supported the changes in the ionizing flux mechanism.

(4) When the data do not include any possible $R$ diagnostics, some works simply assume an $R$ value. The common procedure is to assume a distance where the measured radial velocity ($v$) equals the escape velocity from the supermassive black hole (SMBH): $R=2GM/v^2$, where $M$ is the mass of the SMBH and $G$ is the constant of gravity. This method is popular with the claimed detection of ultra-fast outflows (UFO) in X-ray data \citep[e.g.,][]{Tombesi11}.   

From the discussion above, we conclude that the excited-state trough method is the most reliable and model-independent $R$ determination method for quasar absorption outflows. It is also the only method that can currently probe outflows with  $R$ smaller than a few thousand pc for quasars at redshifts larger than 0.5. 

\subsection{Avoiding Selection Biases When Using  the \siv/\siv* $R$ Determination Method}

\citet{Lucy14} criticized the \siv/\siv* $R$ determination method as suffering from a selection bias, claiming that it cannot be generalized to the whole population of quasar absorption outflows.  They pointed correctly to the fact that, given the critical density for the \siv/\siv* energy levels and the finite quality of observed spectra, there is only a certain range of \ne\ values that can be uniquely determined from the ratio $N$(\siv*)/$N$(\siv). Their given range, 
$4.5\times10^2$ cm$^{-3}$ $<\ne< 1.7 \times 10^5$ cm$^{-3}$, will allow a unique distance determination for the range 50 pc$\ltorder R \ltorder $2000 pc (see \S2.3 here). This in turn will cause a bias toward $R$ values in the range 50--2000 pc, as an accurate \ne\ cannot be measured by the ratio $N$(\siv*)/$N$(\siv) for $\ne> 1.7 \times 10^5$ cm$^3$.

The logical fault with the above argument is that there is important information regarding $R$ even outside the measurable range of \ne\ values. As we detail in this paper, when $N$(\siv*)$_{\mathrm{AOD}}$/$N$(\siv)$_{\mathrm{AOD}}\gtorder1$  we can ascertain only a lower limit for \ne\ and therefore an upper limit of 
$R\ltorder100$ pc. Similarly, when $N$(\siv*)$_{\mathrm{AOD}}$/$N$(\siv)$_{\mathrm{AOD}}<1$ we can ascertain an upper limit for \ne\ and therefore a lower limit of 
$R\gtorder100$ pc. As mentioned above, if the large majority of quasar outflows were situated at $R<1$~pc, then in the large majority of cases, we would have found $N$(\siv*)$_{\mathrm{AOD}}$/$N$(\siv)$_{\mathrm{AOD}}\gtorder1$. Our \siv\ surveys show that this is not the case: roughly 50\% of \siv\ outflows have 
$N$(\siv*)$_{\mathrm{AOD}}$/$N$(\siv)$_{\mathrm{AOD}}<1$ and are therefore at $R>100$~pc.

Our methodolgy specifically avoids a bias toward a certain range of $N$(\siv*)/$N$(\siv) values.  We do so by concentrating on identifying an \siv\ outflow trough and only then establishing whether an associated  \siv* outflow trough exists and at what depth.

\subsection{Implication tfor AGN Wind Models}

Theoretical studies of AGN wind models have a long history. The best-studied models use the AGN accretion disk as the source of the 
wind and accelerate the material with radiation pressure using the central source's luminosity \citep[e.g.][]{Murray95,Proga00,Proga04}. 
The width of the absorption troughs is attributed to the acceleration phase of the winds seen along our line of sight.
In such models for luminous quasars, most of the acceleration is done within several times $10^{16}$~cm ($\sim$0.01 pc) from the central source 
\citep[e.g.][Figure 1]{Higginbottom14}.
Clearly, these accretion disk wind models are hard-pressed to explain absorption outflows situated at 100~pc or more from the central source.

A different set of AGN wind models pertains to lower-luminosity AGNs and attempts to explain their UV troughs  and  warm absorbers (the X-ray absorption manifestation of the wind). In some of these models, the source of the outflowing 
matter is photoionized evaporation from the inner edge of the obscuring torus often
found surrounding an AGN \citep[e.g.,][]{Krolik01}. Other models invoke magnetohydrodynamic (MHD) wind from a clumpy molecular accretion disk to explain
observations of warm absorbing gas at UV and X-ray energies in Seyfert 1 galaxies \citep[e.g.][]{Bottorff00}. Both models create the wind at $R\sim1-10$ ~pc, with wind velocities up to 
$\sim1000$~\kms. The applicability of these models to faster outflows ($\sim10,000$~\kms) in luminous quasars is not well explored in the literature.
 
A different kind of model creates the observed absorption troughs  ``in situ in radiative shocks produced when a
quasar blast wave impacts a moderately dense interstellar clump along the line of sight'' 
at hundreds or thousands of parsecs from the central source \citep{FaucherGiguere12}. 
This model is designed to explain the type of winds we empirically detect on such scales.

\vspace{3mm}
\begin{table}
\caption{Electron Temperature in Kelvin \\ averaged over the \siv\ region.}
\vspace{-3mm}
\begin{center}
\begin{tabular}{r@{~}lrrrr}
&&\multicolumn{4}{c}{$\log\uh$}\\
\multicolumn{1}{l}{Z}&SED&\multicolumn{1}{c}{$-2$}&\multicolumn{1}{c}{$-1$}&\multicolumn{1}{c}{$0$}&\multicolumn{1}{c}{$1$}\\\hline
$1Z_\odot$&MF87&11,000&11,000&12,000&13,000\\
$1Z_\odot$&HE0238&9600&9800&10,000&11,000\\
$1Z_\odot$&UVSoft&11,000&11,000&11,000&12,000\\
$1Z_\odot$&NGC 5548&13,000&12,000&12,000&12,000\\\hline
$4Z_\odot$&MF87&6600&6800&7400&7800\\
$4Z_\odot$&HE0238&6100&6200&6500&7100\\
$4Z_\odot$&UVSoft&7000&6400&6800&7200\\
$4Z_\odot$&NGC 5548&8000&7300&7600&7900\\\hline
\end{tabular}
\end{center}
\label{table3:temp}
\tablecomments{
SED references: \\
- For MF87, HE0238, and HE0238, see \citet{Arav13}, Figure 10 and Table 2. \\
- For NGC5548, see \citet{Steenbrugge2005}, Figure 1.
}
\end{table}

\subsection{The effect of finite Temperature}
\label{sec:Temperature}

Equation (\ref{eqn:ne}) shows that in the high-\ne\ limit, the maximum ratio of 
N(\mbox{\siv*})/N(\mbox{\siv})=2$e^{-\Delta E/kT}$. For \siv, $e^{-\Delta E/kT}=0.8$ at 6000 K, 
so at \necrit,  N(\mbox{\siv*})/N(\mbox{\siv})=0.8. 

Our results in Tables \ref{table_sdss_results} and \ref{table_vlt_results} posit that we treat an outflow as having N(\mbox{\siv*})/N(\mbox{\siv})$<1$
only if the actual ratio is less than 0.8. Therefore, it is important to verify that 
for plasma photoionized by a quasar spectrum, the region where \siv\ exists has $T>6000$ K. 
We used our grid of photoionization models to find the temperature in the \siv\ zone.
We checked the range of $0.01<\uh<10$, which extends more than an order of magnitude on either side of the 
\uh\ solutions for published photoionization analysis of \siv\ outflows \citep{Borguet12b,Borguet13,Chamberlain15a}. 
We did so for three Spectral Energy Distributions (SED) used for quasar outflow studies in the literature and one Seyfert SED. A range of metallicity is $1<Z<4$, where $Z=4$ solar was chosen as  
the upper limit,  since \citet{Borguet12b} showed that it is already too high a metallicity value for an \siv\ outflow.
Table \ref{table3:temp} shows the  temperature over the \siv\ region for the four SED, at four \uh\ values and the chosen metallicity boundaries.
In all cases, $T>6000$ K, thus, our analysis is robust for realistic cases of a quasar-photoionized \siv\ absorber.

\section{Summary}
\label{sec:summary}

In quasar outflow research, arguably the most important
question is: what is the distance of the outflows from the central source?
Most of the fundamental issues in this field are directly tied to the distribution of $R$ values: 

\begin{itemize}

 \item the origin and acceleration mechanism of the outflows.

 \item the influence of the outflow on the formation and evolution of the host galaxy, since the mass, momentum, and kinetic energy fluxes of the outflow depend linearly on $R$ \citep[e.g.][]{Borguet12a}.

 \item the relationship between mass accretion and mass ejection from the environment of the black hole, since the mass flux of the outflow depends linearly on $R$.

\end{itemize}

\begin{table*}
\begin{center}
\caption{Distance from the central source based on an observed \siv~1062\AA\ Trough, \\ and number of outflows in each category}
\begin{tabular}{lr@{\extracolsep{0in}}@{$\,$}c@{$\,$}lr@{\extracolsep{0in}}@{$\,$}c@{$\;$}lrr}
\hline\hline
\siv* 1073\AA\ Trough	&			&\ne	&			&		&$R$	& (pc)				&SDSS		&VLT\\\hline
Undetected		&			&\ne	&$<\neaod<\necrit$	&		&$R$	&$>1000$			 	&	4	& 1 \\
Shallower than \siv	&$\nepi<$		&\ne	&$<\neaod<\necrit$	&$R_{\nepi}>$	&$R$	&$>100$				&	14	& 5 \\
Similar depth to \siv	&$\nepi<$		&\ne	&			&		&$R$	&$<R_{\nepi}$			&	12	& 1 \\
Deeper than \siv	&$\necrit<\neaod<$	&\ne	&			&		&$R$	&$<100$				&	4	& 1 \\ \hline
\end{tabular}
\end{center}
\label{table:distance}
\tablecomments{Definitions: \ne, electron number density; \neaod, value of \ne\ determined from the ratio $N$(\siv*)$_{\mathrm{AOD}}/N$(\siv)$_{\mathrm{AOD}}$(see Fig.~\ref{fig:siv_ap}); 
\necrit, critical density of the \siv/\siv* energy levels, $\ncr=5.6\times10^4$~cm$^{-3}$ at 10,000 K;
\nepi, lower limit on \ne\  from photoionization modeling that give the upper limit to $N$(\siv); and
$R_{\nepi}$, the distance of the outflow when \ne=\nepi, illustrated by the left boundary of the blue and green zones in Fig.~\ref{fig:siv_sim}.
(For elaboration on these quantities, see \S~2.2 and 2.3.)}

\end{table*}

Over the past 15 yr, roughly 20 individual observational studies using excited-state troughs found $10$ pc $\ltorder~R~\ltorder 10,000$ pc, while many theoretical works put the outflows at $R\sim0.01$~pc (see \S~1).  The 20 outflows for which $R$ was determined suffered from selection effects (see \S~1), which did not allow for a simple extrapolation of the results to the majority of quasar outflows.

To overcome these biases, in this paper, we executed the following program:. 

\begin{enumerate}

\item We targeted the \siv\ ion. The fractional abundances of \siv\ and \civ\ peak at  similar values of the ionization
parameter, implying that they arise from the same physical component of the outflow \citep[see Section 3 and Fig.~3 in][]{Dunn12}. Therefore (as discussed in \S~1), the \siv\ troughs give the best distance proxy for the majority of quasar outflows that show \civ\ troughs.

\item In \S~2, we demonstrated that the AOD ratio of the \siv* and \siv\ trough provides robust information about $R$, even while taking into account nonblack saturation that is common in quasar outflow troughs.

\item We then chose two unbiased quasar samples for our \siv\ survey. The first sample included the 1091 brightest SDSS and BOSS quasar spectra, where the spectral regions of possible \siv\ and \siv* outflow troughs are covered (see \S~3.1). The second sample was a blind sample of 13 bright SDSS and BOSS quasars, where the spectral regions of possible \siv\ and \siv* outflow troughs were not covered by the SDSS and BOSS data.  We then observed these quasars using VLT/X-shooter, whose short-wavelength coverage allowed us to observe the spectral region of the \siv\ and \siv* outflow troughs. These observations have much higher spectral resolution and S/N compared to the SDSS and BOSS observations.

\item In \S~4 we identified all the robust cases of \siv\ outflow trough detections and searched for \siv* outflow troughs associated with the same outflow component. We then ran Monte Carlo simulations to quantitatively assess the reliability of the \siv\ trough detection.

\item Following the methodology described in  \S~2 and 3, we separated the \siv\ outflows into different distance groups (Tables 4 and 5). These results are summarized in Table \ref{table:distance}.

\end{enumerate}

{\bf Results} (see Table \ref{table:distance}): In each of the samples we analyze, we find that at least 50\% of quasar outflows have $R$ 
larger than 100-200 pc, and at least 12\% are at distances 
larger than 1000 pc. Higher-quality data hint that a greater fraction  (3/4) of the outflows have $R$ 
larger than 100-200 pc, but our sample size of the VLT/X-shooter outflows is too small (eight outflows) to derive statistically meaningful results.
These $R$ values have profound implications for the study of the origin and 
acceleration mechanism of quasar outflows and their effects on the host galaxy.

\acknowledgments

NA acknowledges support from NSF grant AST 1413319, as well
as NASA STScI grants GO 14242, 14054, and 14176, and NASA ADAP 48020.

Based on observations collected at the European Organisation for Astronomical Research in the Southern Hemisphere
under ESO programs, 091.B-0324, 092.B-0267, 091.B-0324, 091.B-0324, and 090.B-0424.

Funding for SDSS-III has been provided by the Alfred P. Sloan Foundation, the Participating Institutions, the National Science Foundation, and the U.S. Department of Energy Office of Science. The SDSS-III website is http://www.sdss3.org/.

SDSS-III is managed by the Astrophysical Research Consortium for the Participating Institutions of the SDSS-III Collaboration, including the University of Arizona, the Brazilian Participation Group, Brookhaven National Laboratory, Carnegie Mellon University, the University of Florida, the French Participation Group, the German Participation Group, Harvard University, the Instituto de Astrofisica de Canarias, the Michigan State/Notre Dame/JINA Participation Group, Johns Hopkins University, Lawrence Berkeley National Laboratory, the Max Planck Institute for Astrophysics, the Max Planck Institute for Extraterrestrial Physics, New Mexico State University, New York University, Ohio State University, Pennsylvania State University, the University of Portsmouth, Princeton University, the Spanish Participation Group, the University of Tokyo, the University of Utah, Vanderbilt University, the University of Virginia, the University of Washington, and Yale University. 

GL is supported by the National Thousand Young Talents
 Program of China and acknowledges the grant from the National
 Natural Science Foundation of China (No. 11673020 and No. 11421303)
 and the Ministry of Science and Technology of China (National Key
 Program for Science and Technology Research and Development,
 No. 2016YFA0400700).





{\it Facilities:} \facility{VLT} \facility{SDSS}.

\end{document}